\documentclass{article}
\usepackage{graphicx} 
\usepackage{fullpage}
\usepackage{url}
\usepackage{booktabs}
\usepackage{placeins}
\usepackage{multirow}
\usepackage{threeparttable}
\usepackage{pifont} 

\usepackage{xcolor}
\usepackage{amsmath,amssymb}
\usepackage{algorithm}
\usepackage{algpseudocode}
\usepackage{authblk}
\usepackage{hyperref}

\title{High Throughput Meets Low Latency: \\CKKS Functional Bootstrapping on GPUs}
\author[1]{Carlos~Agull{\'o}-Domingo}
\author[2]{Carlo~Pascoe}
\author[1]{{\'O}scar~Vera-L{\'o}pez}
\author[2]{Andreea~Alexandru}
\author[2]{Ahmad~Al~Badawi}
\author[1]{Jos{\'e}~L.~Abell{\'a}n}
\author[2]{Yuriy~Polyakov}
\affil[1]{University of Murcia}
\affil[2]{Duality Technologies}
\date{\today}

\begin{document}

\maketitle

\begin{abstract}

Functional (programmable) bootstrapping extends the bootstrapping operation of Fully Homomorphic Encryption (FHE) to evaluate an arbitrary function while refreshing a ciphertext. This capability was a unique feature of scalar CGGI/DM cryptosystems for many years until it was recently added to the toolset of vectorized FHE schemes, with CKKS being the most promising option both in terms of throughput (amortized time) and functionality. Previous work showed that CKKS functional bootstrapping outperforms CGGI/DM functional bootstrapping in terms of throughput by two to three orders of magnitude, but still loses in latency on CPU by two to three orders of magnitude. We present an optimized GPU implementation of CKKS functional bootstrapping that closes the latency gap. Our implementation supports evaluating functions over both small integers (as in CGGI/DM) and real numbers (unique to CKKS), which we call discrete-CKKS and continuous-CKKS functional bootstrapping, respectively.

Our GPU implementation evaluates discrete-CKKS functional bootstrapping over a ciphertext with 64 encrypted Boolean values in 2.30 ms and with 16,384 encrypted Boolean values in 5.24 ms on RTX 4090, a consumer-grade GPU, whereas the CGGI programmable bootstrapping of one encrypted bit on the same GPU takes 2.16 ms. For 6-bit lookup tables, our discrete-CKKS functional bootstrapping takes 31.4 ms for 32,768 integers vs.\ 15.7 ms for one integer using CGGI. On RTX PRO 6000, our continuous-CKKS functional bootstrapping is 3.5x to 7.4x faster than the best prior GPU implementation on the same GPU, while bootstrapping four to eight times as many values. Our continuous-CKKS functional bootstrapping of the identity function, which corresponds to regular CKKS bootstrapping, takes 4.82 ms for 16,384 slots on RTX PRO 6000, the first reported bootstrapping latency below 5 ms for a CKKS ciphertext with this many slots. We were able to achieve such low latencies via a combination of algorithmic, architectural, and implementation optimizations, many of which were applied to both FIDESlib and OpenFHE. Our continuous-CKKS functional bootstrapping also scales to larger ring dimensions, reaching 28.5 bits of precision with 23 levels remaining at a ring dimension of~$2^{17}$.

\end{abstract}

\section{Introduction}
Bootstrapping is a critical component of any Fully Homomorphic Encryption (FHE) scheme that determines both the functionality and efficiency of FHE in practice~\cite{BP23}. On the one hand, bootstrapping is what gives FHE the fundamental ability to evaluate arbitrary computation circuits by refreshing ``exhausted'' ciphertexts as needed. On the other hand, bootstrapping is the most resource-demanding operation that often limits the practicality of FHE applications. As a result, many efforts have been directed at reducing the latency and/or increasing the throughput of bootstrapping for various FHE schemes, e.g.,~\cite{DM15,CGGI16,JKA+21,OLG+26}.

Two kinds of FHE bootstrapping algorithms are typically distinguished: classical (identity function) bootstrapping and functional/programmable bootstrapping~\cite{BP23}. The distinctive feature of functional bootstrapping is that it supports evaluating a function as part of bootstrapping. The early research, starting with Gentry's monumental work~\cite{Gen09}, focused on the classical bootstrapping algorithms until the authors of the FHEW/Ducas--Micciancio (DM) scheme observed that a Boolean function can be evaluated as part of their bootstrapping method~\cite{DM15}. This observation was subsequently generalized to larger plaintext spaces in the first TFHE/Chillotti--Gama--Georgieva--Izabachène (CGGI) scheme paper~\cite{CGGI16}, and programmable (functional) bootstrapping has become a distinctive feature of CGGI/DM-like schemes. More recently, the functional bootstrapping\footnote{For consistency, we use the term functional bootstrapping throughout the paper; it is equivalent in meaning to programmable bootstrapping.} capability was added to other common FHE schemes, including Brakerski/Fan--Vercauteren (BFV), Brakerski--Gentry--Vaikuntanathan (BGV), and Cheon--Kim--Kim--Song (CKKS)~\cite{LW23,LW24,BKSS24,AKP25}. In the case of CKKS, two variants of functional bootstrapping are further differentiated: discrete~\cite{BKSS24,AKP25,AKPZ26}, which operates on small integers just like DM/CGGI, and continuous~\cite{FEFBT}, which operates on real numbers, similar to regular CKKS. On CPU, discrete-CKKS functional bootstrapping was shown to outperform CGGI/DM functional bootstrapping in terms of throughput by two to three orders of magnitude, but its latency is still two to three orders of magnitude higher~\cite{AKP25}.

Most of the early bootstrapping works focused on algorithmic improvements and on the evaluation of bootstrapping implementations on CPU systems. Despite orders-of-magnitude advances, the latency and throughput of various bootstrapping methods were still not sufficient for many practical applications requiring bootstrapping. This led to multiple hardware acceleration research efforts targeting FPGA~\cite{ACY+23,HGK+25,XAK+26}, GPU~\cite{JKA+21,CKA26,JKS+26}, and ASIC~\cite{FDK+25,KFD+25,KJC+26} systems. Among these different hardware acceleration options, the GPU approach has produced the most remarkable results on real systems, including 7.5 ms in~\cite{JKS+26}, 6.97 ms in~\cite{HEAAN26}, and 8.2 ms in~\cite{OLG+26} for regular CKKS bootstrapping (there are faster simulated results for hypothetical ASIC systems, e.g., 1.38 ms for FAST~\cite{FDK+25} and 0.915 ms for WHET~\cite{KJC+26} for full-slot CKKS bootstrapping at $N=2^{16}$, but no physical chips are commercially available yet). 

\begin{table}[!t]
  \centering
  \caption{Latency [ms] of GPU-based discrete functional bootstrapping of a single ciphertext for plaintext sizes from 1 to 12 bits: the programmable bootstrapping in TFHE-rs~\cite{Zama26} (CGGI), the CKKS bit bootstrapping of HEonGPU~\cite{OS24}, and our discrete-CKKS functional bootstrapping. ``Values'' is the number of values bootstrapped at once. No levels are available after bootstrapping in the CKKS results.}
  \label{tab:sota-discrete}
  \small
  \setlength{\tabcolsep}{5pt}
  \begin{threeparttable}
  \begin{tabular}{lcl rrrrrrr}
    \toprule
    & & & \multicolumn{7}{c}{Plaintext size [bits]} \\
    \cmidrule(lr){4-10}
    Implementation & Values & GPU & 1 & 2 & 4 & 6 & 8 & 10 & 12 \\
    \midrule
    \multirow{2}{*}{TFHE-rs\tnote{a}} & \multirow{2}{*}{1} & RTX 4090 & 2.16 & 2.32 & 4.03 & 15.73 & -- & -- & -- \\
     & & RTX PRO 6000 & 2.46 & 2.63 & 4.55 & 18.67 & -- & -- & -- \\
    \addlinespace
    \multirow{2}{*}{Ours} & \multirow{2}{*}{64} & RTX 4090 & 2.30 & 2.71 & 4.59 & 9.49 & 22.8 & 34.4 & 56.9 \\
     & & RTX PRO 6000 & 2.33 & 2.49 & 3.94 & 7.41 & 15.8 & 23.5 & 39.2 \\
    \addlinespace
    HEonGPU\tnote{b} & 32K & RTX 4090 & 50.72 & -- & -- & -- & -- & -- & -- \\
    \addlinespace
    \multirow{2}{*}{Ours} & \multirow{2}{*}{16K--64K\tnote{c}} & RTX 4090 & 5.24 & 7.40 & 16.6 & 31.4 & 51.5 & 77.6 & -- \\
     & & RTX PRO 6000 & 5.05 & 6.24 & 12.7 & 23.0 & 35.5 & 51.7 & 88.7 \\
    \bottomrule
  \end{tabular}
  \begin{tablenotes}\footnotesize
    \item[a] Run and measured by the authors on the same GPUs.
    \item[b] As reported in the HEonGPU repository for $2^{15}$ bits at ring dimension $2^{16}$ with no levels remaining after bootstrapping.
    \item[c] Fully packed ciphertexts: 16K values for 1 bit, 32K values for 2 to 6 bits, and 64K values for 8 to 12 bits.
  \end{tablenotes}
  \end{threeparttable}

  \vspace{1.5em}

  \centering
  \caption{Latency (``ms'') and precision (``bits'') of GPU-based continuous-CKKS functional bootstrapping of a single ciphertext on RTX PRO 6000, with one level available after bootstrapping: the results of Ouyang et al.~\cite{OLG+26} for 4096 slots ($N = 2^{13}$) and our results ($N = 2^{15}$) for 64, 16,384, and 32,768 slots. The number of slots is the number of real values that are bootstrapped; our case with 32,768 slots corresponds to a fully packed ciphertext, where both the real and the imaginary parts of the 16,384 complex slots are used. The identity function corresponds to regular CKKS bootstrapping.}
  \label{tab:sota-continuous}
  \small
  \begin{tabular}{l rr rr rr rr}
    \toprule
    & \multicolumn{2}{c}{\cite{OLG+26}} & \multicolumn{6}{c}{Ours} \\
    \cmidrule(lr){2-3} \cmidrule(lr){4-9}
    Slots & \multicolumn{2}{c}{4096} & \multicolumn{2}{c}{64} & \multicolumn{2}{c}{16,384} & \multicolumn{2}{c}{32,768} \\
    \cmidrule(lr){2-3} \cmidrule(lr){4-5} \cmidrule(lr){6-7} \cmidrule(lr){8-9}
    Function & ms & bits & ms & bits & ms & bits & ms & bits \\
    \midrule
    Identity & 8.2 & 5.2 & 3.08 & 8.3 & 4.82 & 7.4 & 6.10 & 7.4 \\
    $\tanh$ & 40.4 & 9.2 & 3.57 & 8.6 & 5.49 & 7.4 & 7.01 & 7.4 \\
    Sigmoid & 24.5 & 8.0 & 3.69 & 8.5 & 5.48 & 7.4 & 7.02 & 7.5 \\
    GELU & 40.4 & 6.6 & 3.79 & 9.0 & 5.55 & 7.4 & 7.36 & 7.5 \\
    \bottomrule
  \end{tabular}
\end{table}

The main goal of our work is to advance the state of the art in GPU-accelerated bootstrapping across multiple axes so that FHE can achieve broader adoption in practice. Our key results are compared with the state-of-the-art GPU results in Tables~\ref{tab:sota-discrete} and~\ref{tab:sota-continuous}. First, we provide the first GPU implementation of discrete-CKKS functional bootstrapping for plaintext moduli larger than two (Table~\ref{tab:sota-discrete}), which reduces the latency by more than two orders of magnitude compared to the state-of-the-art CPU results~\cite{OpenFHEPerf}; in the Boolean case, a fully packed ciphertext is bootstrapped in 5.24 ms on RTX 4090, which is 9.7x lower than the latency of the bit bootstrapping in~\cite{OS24}. Second, we report a 3.5x to 7.4x lower latency of GPU-based continuous-CKKS functional bootstrapping compared to~\cite{OLG+26}, while bootstrapping four to eight times as many values (Table~\ref{tab:sota-continuous}). Third, our work is the first to achieve a latency of less than 5 ms for regular CKKS bootstrapping (the identity function): 4.82 ms for 16,384 slots, compared to 8.2 ms for 4096 slots in~\cite{OLG+26}. Fourth, the latency of our discrete-CKKS functional bootstrapping is lower than or comparable to that of the state-of-the-art CGGI implementation for a small number of slots (e.g., 7.41 ms for 64 six-bit values vs.\ 18.67 ms for one value on RTX PRO 6000) and only moderately higher (by at most a factor of about four) for fully packed ciphertexts, where the throughput of our CKKS implementation is three to four orders of magnitude higher (Table~\ref{tab:sota-discrete}). We intend to make our implementation available in later releases of OpenFHE and FIDESlib.

\subsection{Technical Overview}
Discrete-CKKS and continuous-CKKS functional bootstrapping extend the workflow used for regular CKKS bootstrapping, which typically includes the slot-to-coefficient (\textsc{StC}) linear transformation, raising the modulus (\textsc{ModRaise}), the coefficient-to-slot (\textsc{CtS}) linear transformation, and approximate modular reduction (\textsc{EvalMod}) steps. \textsc{EvalMod} usually requires evaluating a large-degree polynomial that approximates a trigonometric function, e.g., sine, using a Chebyshev series interpolation, which is often evaluated using the Paterson--Stockmeyer (PS) algorithm. In the case of CKKS functional bootstrapping, another step is added to evaluate a trigonometric interpolation corresponding to either a lookup table (LUT) in the discrete case or a relatively smooth function in the continuous case. The trigonometric interpolation is represented as a power series, e.g., over a complex exponential function, the degree of which grows with the plaintext modulus or the target precision. For better efficiency, the power series is also evaluated using the PS method. The above implies that the PS algorithm for evaluating a large-degree polynomial becomes the bottleneck of CKKS functional bootstrapping, especially when higher precision is sought.

To achieve low latency for CKKS functional bootstrapping, we started by designing a highly efficient PS core routine that is tailored for evaluating a large-degree polynomial on both CPUs and GPUs. Our design is governed by three principles. First, the linear combinations at the leaves of the PS recursion are produced directly at the level and scaling factor required by the operations that consume them, which eliminates the level adjustments and ciphertext copies of prior implementations. Second, a rescaling or relinearization is performed only when the next non-scalar multiplication requires it, so that no value is rescaled twice and the number of rescalings grows as the square root of the polynomial degree rather than linearly. Third, all leaves are evaluated in a single batched pass, and the remaining non-scalar multiplications are grouped by their height in the recursion tree, which exposes the parallelism of the algorithm to both multi-threaded CPUs and GPUs. The same routine is used for the Chebyshev and power series. For polynomial degrees from 27 to 4031, the CPU implementation of the PS core routine is 1.8x to 10.9x faster than the PS implementation in OpenFHE v1.6.1~\cite{ABB+22,OpenFHE161} in the single-threaded mode, with the speedup growing with the degree, and 3.4x to 8.9x faster with 36 threads.

To support smaller FHE parameters, we revisited the bootstrapping designs in OpenFHE, focusing on reducing the multiplicative depth required for bootstrapping circuits. First, we eliminated the level consumed right after \textsc{ModRaise} by using an approach inspired by the OverModRaise techniques in~\cite{KCY25}, which was especially challenging for the case of sparse packing and encapsulated sparse secrets (the original paper does not consider sparse packing and only briefly discusses encapsulation, with a placement of the key switching that loses precision). Second, for continuous functional bootstrapping, we eliminated another level needed for the adjustment/correction before \textsc{ModRaise} by embedding it in the preceding \textsc{StC} matrices. Third, we saved up to two more levels in continuous functional bootstrapping for small scaling factors, where coarser approximations become admissible. Fourth, we introduced a new encapsulation mode where uniform ternary secrets are used for the main computation, which increases the maximum ciphertext modulus supported for a given ring dimension. Finally, we fused key switching (relinearization) with the subsequent rescaling, which reduces the computational complexity of the bootstrapping algorithms. As a result of these optimizations and the new PS core routine, our CPU implementation of discrete functional bootstrapping is 3.3x to 5.1x faster than OpenFHE v1.6.1~\cite{OpenFHEPerf} for the same configurations with plaintext sizes from 1 to 8 bits, and 3.9x to 4.8x faster when the new encapsulation mode is used.

All of the above improvements apply to both CPU and GPU. After that, we switched our attention to GPU-specific optimizations in FIDESlib 2.2.0~\cite{AVG+25}. First, since functional bootstrapping can often use scaling factors of at most 30 bits, we added support for 32-bit limbs, which nearly halves the memory footprint and makes the number-theoretic transform faster than with the emulated 64-bit arithmetic. Second, we implemented the switch-back and partial sum of our OverModRaise variant with two fused kernels around a single modulus reduction. Third, we captured the whole bootstrapping procedure as a CUDA graph with a static memory pool and pruned dependencies, which reduces the latency by up to 2.3x when it is dominated by kernel launches. We also use the sparse plaintext compression of~\cite{JKS+26}, which reduces the memory footprint of the linear transformations by 2.9x for fully packed ciphertexts and by much more for sparse packing.

\subsection{Related Work}

The work most closely related to ours is the recent sub-10 ms CKKS bootstrapping result of Ouyang et al.~\cite{OLG+26}. Instead of approximating modular reduction with a large-degree polynomial in \textsc{EvalMod}, as is done in our work and in most other CKKS bootstrapping implementations, the authors build on the SHIP algorithm~\cite{CHKS25}, which computes the roots of unity $\omega^{m_i}$ directly via blind rotations driven by a sparse secret key (Hamming weight 31) and combines them using a product tree. Their implementation bootstraps a single ciphertext in 8.2 ms on RTX PRO 6000 at $N = 2^{13}$ with approximately 5 bits of precision and one remaining level, and the same machinery yields continuous functional bootstrapping in 24.5--40.4 ms at 6.6--9.2 bits of precision ($N = 2^{13}$) or in 248--382 ms at 23.3--24.0 bits of precision ($N = 2^{14}$). Our work differs in several respects. First, we retain the standard \textsc{EvalMod}-based workflow with dense (encapsulated) secrets, so that our optimizations apply directly to widely used libraries such as OpenFHE and FIDESlib. Second, Ouyang et al. only evaluate continuous functional bootstrapping for smooth functions and report no results for discrete functional bootstrapping over small integers. Third, the two approaches scale differently with the ring dimension, and hence with the precision and the number of remaining levels, as both require a larger modulus and therefore a larger $N$ for the same security level. The blind rotation of~\cite{OLG+26} performs about $2N$ plaintext-ciphertext multiplications and uses a key of about $N$ ciphertexts, each over the raised modulus with $\ell$ residue number system (RNS) limbs, so both its computational cost and its key size grow as $\Theta(N^2 \ell)$. In the standard workflow, the linear transformations with a level budget of $b$ require $O(b N^{1/b})$ plaintext-ciphertext multiplications and $O(b N^{1/(2b)})$ key switchings, and the number of non-scalar multiplications of the polynomial evaluation does not depend on $N$, so the cost grows as $\tilde{O}(N^{1+1/b} \ell)$. This is reflected in the results of~\cite{OLG+26}, where doubling the ring dimension increases the latency of regular bootstrapping from 8.2 ms to 44 ms and the latency of functional bootstrapping by an order of magnitude, whereas our approach achieves 4.82 ms for 16,384 slots at $N = 2^{15}$ (Table~\ref{tab:sota-continuous}) and scales to 23--24 remaining levels and 28.5 bits of precision at $N = 2^{17}$ (Section~\ref{sec:results-continuous}).

HEonGPU~\cite{OS24} is an open-source GPU library for BFV, CKKS, and TFHE that supports several variants of CKKS bootstrapping, including a bit bootstrapping for binary plaintexts. On RTX 4090, it reports a latency of 50.72 ms for bootstrapping $2^{15}$ bits at $N = 2^{16}$ with no levels remaining after bootstrapping. To the best of our knowledge, this Boolean case is the only instance of discrete-CKKS functional bootstrapping that was previously implemented on a GPU. Our implementation supports arbitrary lookup tables over larger plaintext moduli, and in the Boolean case it bootstraps a fully packed ciphertext of $2^{14}$ bits in 5.24 ms on the same GPU (Table~\ref{tab:sota-discrete}), i.e., at a 9.7x lower latency and a 4.8x lower amortized cost per bit.

Cerium~\cite{JKS+26} and the closed-source HEAAN library by CryptoLab~\cite{HEAAN26} report the lowest GPU latencies for regular CKKS bootstrapping at high precision. Cerium is a multi-GPU framework that combines a domain-specific language, an optimizing compiler that automatically fuses limb-level kernels, and a runtime system, with the goal of running large encrypted inference workloads; for $N = 2^{16}$, approximately 20 bits of precision, and 18 remaining levels, it reports a latency of 7.5 ms on a system with eight B200 GPUs and 14.5 ms on a single B200. HEAAN advertises a latency of 6.97 ms on RTX 5090 for 32K slots and 19 bits of precision, but no implementation details behind this number have been disclosed, so it cannot be independently verified or compared under matched conditions. Neither work considers functional bootstrapping. Our latency of less than 5 ms for regular CKKS bootstrapping, which we obtain as the identity function in continuous functional bootstrapping, is achieved in a different regime, namely at low precision with one remaining level, where much smaller parameters can be used; in the high-precision regime targeted by these works, our focus is on functional bootstrapping rather than on the lowest latency of regular bootstrapping.

There are also several recent works that reduce the latency of (functional) CKKS/BFV bootstrapping at the expense of decreasing the number of slots~\cite{SWIFT,ZLW26,park2026} (beyond the trivial linear transformation efficiency improvements for sparsely packed ciphertexts). In this case, multiple polynomial coefficients are used to represent one integer (or real number), and then this ``block'' representation is employed to speed up the polynomial evaluation. Our work focuses on simultaneously achieving high throughput and low latency, as this is often needed in ML/AI and large-scale database applications. Therefore, we consider the full-packing case as our main target. Some of the throughput-reducing latency optimizations can potentially be applied to our design as well if the goal is to achieve even lower latency, but this exploration is left for future work.

For CGGI/DM-like schemes, the most relevant results are the GPU benchmarks of the TFHE-rs library by Zama~\cite{Zama25,Zama26}. TFHE-rs reports programmable bootstrapping latencies on H100 of 0.796 ms for Boolean ciphertexts and 0.874, 0.971, and 9.9 ms for 2-, 4-, and 6-bit plaintexts, respectively. To compare the two approaches on the same hardware, we also ran TFHE-rs on our GPUs: the corresponding latencies are 2.16, 2.32, 4.03, and 15.73 ms on RTX 4090 and 2.46, 2.63, 4.55, and 18.67 ms on RTX PRO 6000 (Table~\ref{tab:sota-discrete}), and between 0.92 and 5.01 ms on the datacenter B200 GPU (Section~\ref{sec:results-discrete}). These latencies are for a single LWE ciphertext encrypting one value, and larger plaintext moduli are considerably more expensive, as illustrated by the jump from 4 to 6 bits. Our discrete-CKKS functional bootstrapping instead refreshes tens of thousands of integers in a single bootstrapping operation. On RTX 4090 and RTX PRO 6000, its latency for 64 values is lower than or comparable to the TFHE-rs latency for one value, and its latency for a fully packed ciphertext is at most about four times higher, which corresponds to a throughput that is three to four orders of magnitude higher; on B200, TFHE-rs has the lower latency for a single value.

BOLT-FHE~\cite{CZF+26} is a recent throughput-oriented GPU implementation of DM/CGGI bootstrapping that keeps the entire blind-rotation accumulator chain on chip (in registers and shared memory) within a single fused kernel, and supports gadget decomposition, modulus raising, and common-mask packing in a unified kernel template. On RTX 4090, BOLT-FHE reaches 40,166 gate bootstrappings per second at 128-bit security (modulus raising with common-mask packing of five ciphertexts), and 11,423 bootstrappings per second for the OpenFHE STD128 gadget-decomposition parameters. These numbers are obtained for batches of 16,384 bootstrapping instances and are reported only as throughput; the latency of bootstrapping a single ciphertext is not given. Since the throughput of BOLT-FHE is measured in bootstrapped values per second (one Boolean value per bootstrapping), it can be directly compared with our discrete-CKKS functional bootstrapping. On the same GPU, our implementation bootstraps a fully packed ciphertext of 16,384 Boolean values in 5.24 ms (Table~\ref{tab:sota-discrete}), which corresponds to 3.13 million values per second, i.e., a 78x higher throughput than the peak throughput of BOLT-FHE, without any batching across ciphertexts. For larger plaintext sizes, our throughput on RTX 4090 remains above one million values per second for up to 8 bits (e.g., 65,536 8-bit values are bootstrapped in 51.5 ms).

\paragraph{Concurrent work.}
TUTTI~\cite{KG26} is a concurrent work that targets low-latency CKKS bootstrapping on multi-threaded CPUs. It builds on SPRU~\cite{CK27}, which, similarly to SHIP~\cite{CHKS25}, evaluates the decryption equation on the complex roots of unity under a sparse secret instead of approximating the modular reduction by a polynomial, and extends it to more slots and to a higher degree of parallelism. TUTTI focuses on bootstrapping one or a small number of slots: it bootstraps a single value in 4.6 ms on 32 threads at 6.5 bits of precision with one remaining level, and its cost grows with the number of slots (e.g., 4096 values take 46 ms on 192 threads). Functional bootstrapping is not considered in that work: TUTTI only refreshes fixed-point values, and its authors suggest switching to CGGI for the evaluation of arbitrary functions. We do not provide a detailed comparison, as this work is concurrent and targets CPUs. In contrast to TUTTI, we focus on functional bootstrapping and on a design that achieves low latency and high throughput at the same time.

\subsection{Organization}
The rest of the paper is organized as follows. Section~\ref{sec:prelim} provides the necessary background on CKKS and on its discrete and continuous functional bootstrapping. Section~\ref{sec:ps} presents our algorithm for the evaluation of large-degree polynomials. Section~\ref{sec:algopt} describes the algorithmic optimizations that reduce the multiplicative depth and the computational cost of bootstrapping. Section~\ref{sec:gpu} describes the design of our GPU implementation in FIDESlib. Section~\ref{sec:results} presents the experimental results for polynomial evaluation and for discrete and continuous functional bootstrapping. Section~\ref{sec:conclusions} concludes the paper.

\section{Preliminaries}
\label{sec:prelim}

\subsection{Notation}

All logarithms are in base~2 unless indicated otherwise. Vectors and ring elements are written in bold, and a polynomial $\mathbf{m}(X)$ is identified with the vector $\mathbf{m}$ of its coefficients. For $x \in \mathbb{R}$, $\lfloor x \rceil$ denotes rounding to the closest integer; for a vector or polynomial, rounding and modular reduction are applied coefficient-wise. We write $[p] = \{0, 1, \ldots, p-1\}$, use $[p]$ as the set of representatives of $\mathbb{Z}_p$, and denote by $\omega_p = \exp(2\pi i/p)$ a primitive $p$-th root of unity. For a function $f: \mathbb{Z}_p \to \mathbb{Z}_p$ and a vector $\mathbf{z} \in \mathbb{Z}_p^n$, $f(\mathbf{z})$ denotes the entry-wise application of $f$; the same convention is used for $\omega_p^{\mathbf{z}}$. Let $N$ be a power of two (for efficiency) and $R = \mathbb{Z}[X]/(X^N+1)$; for an integer $q > 1$, $R_q = R/qR = \mathbb{Z}_q[X]/(X^N+1)$. For a polynomial $\mathbf{a}$, $\|\mathbf{a}\|$ denotes the infinity norm of its coefficient vector. Let $M = 2N$, $\zeta = \exp(2\pi i/M)$, and $\mathbb{Z}_M^* = \{x \in \mathbb{Z}_M : \gcd(x, M) = 1\}$. For a power of two $n \le N/2$, the canonical embedding $\tau'_n : \mathbb{R}[X]/(X^N+1) \to \mathbb{C}^n$ evaluates a polynomial at the $n$ roots $\zeta^{5^j}$, $j \in [n]$, and is the map used for CKKS encoding and decoding~\cite{CKKS17}.

\subsection{RLWE Encryption}

The secret-key ring learning with errors (RLWE) encryption scheme~\cite{LPR10} is parameterized by a plaintext modulus $p$, a ciphertext modulus $q'$, the ring dimension $N$, and two small distributions $\chi_{\mathrm{key}}$ and $\chi_{\mathrm{err}}$ over $R$. A vector of integers from $\mathbb{Z}_p$ is encoded as a polynomial $\mathbf{m} \in R_p$ whose $k$-th coefficient is the $k$-th entry of the vector (coefficient encoding).
\begin{itemize}
  \item $\mathsf{Encrypt}(\mathbf{m}, \mathbf{s})$: for $\mathbf{s} \leftarrow \chi_{\mathrm{key}}$, uniform $\mathbf{a} \leftarrow R_{q'}$, and $\mathbf{e} \leftarrow \chi_{\mathrm{err}}$, output $(\mathbf{b}, \mathbf{a}) \in R_{q'}^2$ with $\mathbf{b} = -\mathbf{a} \cdot \mathbf{s} + (q'/p) \cdot \mathbf{m} + \mathbf{e}$.
  \item $\mathsf{Decrypt}((\mathbf{b}, \mathbf{a}), \mathbf{s})$: output $\left\lfloor (p/q') \cdot (\mathbf{b} + \mathbf{a} \cdot \mathbf{s}) \right\rceil \bmod p$.
  \item $\mathsf{ModSwitch}((\mathbf{b}, \mathbf{a}), q', Q')$: output $\left(\lfloor (Q'/q') \cdot \mathbf{b} \rceil,\, \lfloor (Q'/q') \cdot \mathbf{a} \rceil\right) \in R_{Q'}^2$.
\end{itemize}
Decryption is exact as long as $\|\mathbf{e}\| < q'/(2p)$. This encryption coincides with BFV in coefficient encoding; the public-key variant is obtained in the standard way. In this work, the RLWE scheme is only used for encryption and decryption of discrete data, and all homomorphic computations are performed in CKKS.

\subsection{CKKS Encryption}

We consider the RNS variant~\cite{CHK+19} of the CKKS scheme~\cite{CKKS17}, together with the reduced-approximation-error modifications of~\cite{KPP22}, which is the variant implemented in OpenFHE. To distinguish RLWE parameters from CKKS parameters, the RLWE moduli are denoted with a prime, e.g., $q'$, whereas the CKKS moduli are unprimed. We select $L+1$ primes $q_0, q_1, \ldots, q_L$ with $q_i \equiv 1 \bmod 2N$ and define $Q_\ell = \prod_{i=0}^{\ell} q_i$ for $\ell \in [L+1]$, where $L$ is the maximum multiplicative depth; elements of $\mathbb{Z}_{Q_\ell}$ are represented by their residues modulo $q_0, \ldots, q_\ell$. The scaling factor is $\Delta = 2^\rho$ with $\Delta < q_0$. A message is a vector $\mathbf{z} \in \mathbb{C}^n$ with $n \mid N/2$ ($n = N/2$ is called full packing, $n < N/2$ sparse packing), and a plaintext is a polynomial in~$R$. A ciphertext at level $\ell$ is a pair $\mathbf{ct} = (\mathbf{c}_0, \mathbf{c}_1) \in R_{Q_\ell}^2$.
\begin{itemize}
  \item $\mathsf{KeyGen}$: sample $\mathbf{s} \leftarrow \chi'_{\mathrm{key}}$, uniform $\mathbf{a} \leftarrow R_{Q_L}$, and $\mathbf{e} \leftarrow \chi_{\mathrm{err}}$; set $\mathsf{sk} = (1, \mathbf{s})$ and $\mathsf{pk} = (\mathbf{b}, \mathbf{a})$ with $\mathbf{b} = -\mathbf{a} \cdot \mathbf{s} + \mathbf{e} \bmod Q_L$.
  \item $\mathsf{Encode}(\mathbf{z}, \Delta)$: for $\mathbf{z} \in \mathbb{C}^n$, output $\mathbf{m} = \lfloor \tau'^{-1}_n(\Delta \cdot \mathbf{z}) \rceil \in R$ (slot encoding).
  \item $\mathsf{Decode}(\mathbf{m}, \Delta)$: for $\mathbf{m} \in R$, output $\mathbf{z} = \tau'_n(\mathbf{m}/\Delta) \in \mathbb{C}^n$.
  \item $\mathsf{Encrypt}(\mathbf{m}, \mathsf{pk})$: sample $\mathbf{v}$ from a small distribution $\chi_{\mathrm{enc}}$ over $R$ and $\mathbf{e}_0, \mathbf{e}_1 \leftarrow \chi_{\mathrm{err}}$; output $\mathbf{ct} = \mathbf{v} \cdot \mathsf{pk} + (\mathbf{m} + \mathbf{e}_0, \mathbf{e}_1) \bmod Q_L$.
  \item $\mathsf{Decrypt}(\mathbf{ct}, \mathsf{sk})$: for $\mathbf{ct} = (\mathbf{c}_0, \mathbf{c}_1) \in R_{Q_\ell}^2$, output $\tilde{\mathbf{m}} = \mathbf{c}_0 + \mathbf{c}_1 \cdot \mathbf{s} \bmod Q_\ell$.
\end{itemize}
CKKS supports homomorphic addition, multiplication, and cyclic rotation of the slots; ciphertext-ciphertext multiplication and rotation require a key-switching operation. A multiplication increases the scaling factor from $\Delta$ to $\Delta^2$, which is restored by a rescaling operation that divides the ciphertext by $q_\ell$ and moves it from level $\ell$ to $\ell-1$. We say that a ciphertext with scaling factor $\Delta$ has noise degree one, and that an unrescaled product, with scaling factor $\Delta^2$, has noise degree two. After $L$ multiplicative levels, a ciphertext in $R_{q_0}^2$ no longer supports multiplications and has to be bootstrapped. All large-degree polynomial evaluations in this work, both in regular and functional bootstrapping, use the Paterson--Stockmeyer (PS) algorithm~\cite{PS73}, which evaluates a degree-$d$ polynomial using $\sqrt{2d} + O(\log d)$ non-scalar multiplications.

The secret distribution $\chi'_{\mathrm{key}}$ is either the uniform ternary distribution or a sparse ternary distribution with $h$ non-zero coefficients. Sparse secrets lower the cost of bootstrapping, as the integer overflow introduced by modulus raising (see below) is bounded by $h$, but also decrease the security level. We use sparse-secret encapsulation~\cite{BTPH22}, where the main computation is performed under a secret with a larger Hamming weight (or a uniform ternary secret) and the ciphertext is switched to a sparser secret only immediately before bootstrapping (and back after it), which retains the efficiency of sparse secrets and makes the bootstrapping failure probability negligible without increasing the ring dimension.

\paragraph{CKKS bootstrapping.}
Bootstrapping takes a ciphertext in $R_{q_0}^2$ encrypting a slot-encoded plaintext and outputs a ciphertext in $R_{Q_{L_0}}^2$ with $L_0 > 0$ that encrypts (approximately) the same message. We present here the variant of the original procedure~\cite{CHK+18} in which the slot-to-coefficient transformation is performed first:
\begin{enumerate}
  \item $\textsc{StC}$: homomorphically evaluate the decoding map $\tau'_n$ so that the message is moved from the slots to the coefficients of the plaintext, i.e., the result encrypts $\mathbf{m}(X)$ with $\mathbf{m} = \lfloor \Delta \cdot \mathbf{z} \rceil$ in coefficient encoding.
  \item $\textsc{ModRaise}$: reinterpret the ciphertext in $R_{q_0}^2$ as a ciphertext in $R_{Q_L}^2$. The new ciphertext decrypts to $\mathbf{t}(X) = \mathbf{m}(X) + q_0 \cdot \mathbf{I}(X)$, where $\mathbf{I}$ is an integer polynomial with $\|\mathbf{I}\| < K$ and $K$ depends on the secret distribution.
  \item $\textsc{CtS}$: homomorphically evaluate the encoding map $\tau'^{-1}_n$ so that each coefficient $t_i$ of $\mathbf{t}(X)$ is placed in a separate slot (two ciphertexts for full packing, one for sparse packing).
  \item $\textsc{EvalMod}$: homomorphically evaluate the modular reduction $t_i \mapsto [t_i]_{q_0} = m_i$ in every slot. As $|m_i| \ll q_0$, it is approximated by the scaled sine $\frac{q_0}{2\pi}\sin\left(\frac{2\pi t_i}{q_0}\right)$, which is in turn approximated by a polynomial, typically a Chebyshev interpolant of a cosine followed by double-angle formulas~\cite{HK20,BMTH21}.
\end{enumerate}
\textsc{StC} and \textsc{CtS} are linear transformations implemented as plaintext-ciphertext matrix products using the baby-step giant-step algorithm; they consume $L_{\mathrm{StC}}$ and $L_{\mathrm{CtS}}$ levels, respectively, and \textsc{EvalMod} consumes $L_{\mathrm{mod}}$ levels. The number of levels remaining after bootstrapping is $L_0 = L - L_{\mathrm{StC}} - L_{\mathrm{CtS}} - L_{\mathrm{mod}}$. Since the scaled sine approximates $[t]_{q_0}$ only in a small neighborhood of the multiples of $q_0$, the message is scaled down by a factor $\delta$ (typically $2^{-8}$ to $2^{-10}$) before bootstrapping, which trades a few bits of plaintext space for a lower-degree \textsc{EvalMod} polynomial.

\subsection{Discrete-CKKS Functional Bootstrapping}
\label{sec:prelim-discrete}

Discrete CKKS~\cite{BKSS24,AKP25} uses CKKS as a computation engine for data in $\mathbb{Z}_p$, where $p$ is a few bits long, and evaluates an arbitrary function $f: \mathbb{Z}_p \to \mathbb{Z}_p$, represented as a lookup table $\mathrm{LUT}(f)$, as part of bootstrapping. We follow the hybrid RLWE/CKKS formulation of~\cite{AKP25}: data is encrypted and decrypted using the RLWE scheme, which achieves exact correctness, and converted to CKKS only for functional bootstrapping. Given an RLWE ciphertext $\mathbf{ct} \in R_{q'}^2$ encrypting $\mathbf{m} \in R_p$, the procedure consists of the following steps.
\begin{enumerate}
  \item $\textsc{ModSwitch}$: switch the modulus from $q'$ to $q_0$. The result is interpreted as a CKKS ciphertext in $R_{q_0}^2$ encrypting $\mathbf{m}/p$ in coefficient encoding with scaling factor $\Delta = q_0$.
  \item $\textsc{ModRaise}$ and $\textsc{CtS}$: as in regular bootstrapping. After $\textsc{CtS}$, the $i$-th slot contains $\Delta \cdot (m_i/p + I_i + v_i)$, where $I_i$ is the integer overflow introduced by $\textsc{ModRaise}$ and $v_i$ is the (normalized) noise of the input ciphertext together with the noise introduced by $\textsc{CtS}$.
  \item $\textsc{EvalExp}$: homomorphically evaluate the complex exponential $x \mapsto \exp(2\pi i x)$, so that the $i$-th slot becomes $\Delta \cdot \omega_p^{m_i}$ (up to noise). This step removes the overflow $I_i$ and plays the role of \textsc{EvalMod}. The exponential is approximated by a Chebyshev interpolant on the interval $[-K, K]$ and refined using repeated squaring.
  \item $\textsc{EvalLUT}_f$: homomorphically evaluate a trigonometric interpolation of $f$ at the roots of unity. In the first-order trigonometric Hermite interpolation of~\cite{AKP25}, one evaluates the polynomial $T_f \in \mathbb{C}[z]$ of degree $p-1$ in $z = \omega_p^{m_i}$ satisfying
  \begin{equation*}
    \operatorname{Re}\left(T_f(\omega_p^k)\right) = f(k)/p \quad \text{and} \quad \operatorname{Re}\left(T'_f(\omega_p^k)\right) = 0 \quad \text{for all } k \in [p],
  \end{equation*}
  and takes the real part. The vanishing first derivative removes the first-order term of the input noise, so that the output noise is quadratically smaller than the input noise (noise cleaning); higher-order Hermite interpolation yields a stronger cleaning at the cost of a higher degree~\cite{AKP25,AKPZ26}.\footnote{The original method~\cite{AKP25} introduced intermediate imaginary-component noise that limited the overall noise reduction, but this extra noise cost was eliminated in~\cite{AKPZ26} by setting the imaginary-component derivatives to zero while keeping the additional computational overhead to a minimum via various algorithmic optimizations. In this work, we use the implementation of~\cite{AKP25} in OpenFHE.} The $i$-th slot is now $\Delta \cdot (f(m_i)/p + v'_i)$ with $|v'_i| \ll |v_i|$. 

  \item $\textsc{StC}$ and $\textsc{ModSwitch}$: move the result back to coefficient encoding, scaling by $Q/(\Delta p)$, where $Q$ is the CKKS modulus after bootstrapping, and switch the modulus to $Q'$ to obtain an RLWE ciphertext encrypting $f(\mathbf{m}) \in R_p$ with ciphertext modulus $Q'$.
\end{enumerate}
Decryption of the final RLWE ciphertext is exact as long as $\|\mathbf{v}'\| < 1/(2p)$. Additional leveled CKKS computations may be performed between the last two steps. Compared to regular bootstrapping, the levels consumed by \textsc{EvalMod} are replaced by those of \textsc{EvalExp} and \textsc{EvalLUT}; the latter dominate the cost for larger $p$, as the degree of $T_f$ grows linearly in $p$ and the interpolation is evaluated as a power series in $z$ using the PS algorithm. Several LUTs can be evaluated on the same input by reusing the powers of $z$ (multi-value bootstrapping)~\cite{AKP25,DAC+25}.

\subsection{Continuous-CKKS Functional Bootstrapping via Fourier Extension}
\label{sec:prelim-continuous}

Continuous-CKKS functional bootstrapping evaluates a real-valued function $f_T$, e.g., a sigmoid or GELU, on real-valued inputs during bootstrapping. We follow the Fourier-extension approach of~\cite{FEFBT}. After $\textsc{StC}$, $\textsc{ModRaise}$, and $\textsc{CtS}$, each slot contains $t_i = m_i + q_0 I_i$ with $m_i \in [-\delta q_0/2, \delta q_0/2]$ (recall $\delta\in (0,1]$ is the fraction of the period $[-q_0/2, q_0/2]$ occupied by the message), so the function to be evaluated is $x \mapsto f_T([x]_{q_0})$, a $q_0$-periodic function. After rescaling the slots by $1/q_0$, the task reduces to evaluating a $1$-periodic function $\hat{f}$ that agrees with $f_T$ on $[-\delta/2, \delta/2]$ by a truncated Fourier series
\begin{equation}
  \label{eq:fourier}
  S_d(x) = \sum_{k=-d}^{d} c_k \exp(2\pi i k x) = c_0 + 2\operatorname{Re}\left(\sum_{k=1}^{d} c_k z^k\right), \qquad z = \exp(2\pi i x),
\end{equation}
where the second equality holds for real-valued $\hat{f}$. The number of terms $d$ needed for a given precision is governed by the smoothness of $\hat{f}$: the Fourier coefficients of a $C^\kappa$ periodic function decay as $O(k^{-\kappa-1})$. Since a direct periodic extension of $f_T$ is in general discontinuous at the endpoints of $[-\delta/2, \delta/2]$,~\cite{FEFBT} instead uses a $\kappa$-th order Fourier extension, which appends a bridge polynomial of degree $2\kappa+1$ on $[\delta/2, \delta/2 + \delta_B]$, where $\delta_B$ is the width of the bridge interval, matching the first $\kappa$ derivatives of $f_T$ at both endpoints, and then extends periodically with period $\delta + \delta_B \le 1$; the resulting $\hat{f}$ is $C^\kappa$, and the unused part of the period absorbs the overflow $I_i$ exactly as in regular bootstrapping.

Homomorphically, \textsc{EvalMod} is replaced by $\textsc{EvalSeries}$, which consists of $\textsc{EvalExp}$ (computing $z = \exp(2\pi i t_i/q_0)$ as in Section~\ref{sec:prelim-discrete}) followed by the evaluation of the degree-$d$ power series in~\eqref{eq:fourier} using the PS algorithm. The degree $d$ grows with the target precision and with $1/\delta$, so, as in the discrete case, the PS evaluation of a large-degree polynomial is the dominant cost. Regular CKKS bootstrapping is the special case $f_T(x) = x$: the identity extended with period $q_0$ is exactly the modular reduction $[x]_{q_0}$ approximated in \textsc{EvalMod}, and the scaled-sine approach corresponds to using the single-term series $S_1$ together with a small~$\delta$. In this work, we instead bootstrap with the Fourier extension of the identity function from~\cite{FEFBT}, with $\delta = 1/2$. The scaled-sine approximation in regular CKKS bootstrapping is accurate only in a small neighborhood of the multiples of $q_0$, which forces a large gap between $q_0$ and the message (the small $\delta$) and wastes many bits of plaintext space (often close to ten bits). The Fourier extension of the identity has no such restriction, as it approximates the modular reduction accurately over half of the period, so the gap between $q_0$ and the message can be made small. This yields a higher output precision for the same $q_0$ and, conversely, allows a smaller $q_0$ and scaling factor $\Delta$, and hence smaller lattice parameters, for a given target precision~\cite{FEFBT}. We therefore use the Fourier-extension method uniformly for both regular bootstrapping and non-trivial $f_T$.

\section{Efficient Polynomial Evaluation}
\label{sec:ps}

All three bootstrapping variants of Section~\ref{sec:prelim} typically spend most of their time evaluating large-degree polynomials: a Chebyshev interpolant of the complex exponential in \textsc{EvalExp}, the degree-$(p-1)$ trigonometric Hermite interpolation in \textsc{EvalLUT}, and the degree-$d$ Fourier series in \textsc{EvalSeries}. This section describes our evaluation algorithm for such polynomials. We first recall the Paterson--Stockmeyer (PS) algorithm as implemented in OpenFHE v1.6.1~\cite{OpenFHE161}, and then present the algorithmic improvements that reduce its number of relinearizations, rescalings, level adjustments, and rounding operations, and expose the parallelism that our GPU implementation (Section~\ref{sec:gpu}) exploits.

\subsection{Paterson--Stockmeyer Evaluation in the Power and Chebyshev Bases}
\label{sec:ps-base}

Let $f$ be a polynomial of degree $d$ given by its coefficients in a basis $\{B_i\}_{i \ge 0}$, which is either the power basis $B_i(x) = x^i$ or the Chebyshev basis $B_i(x) = T_i(x)$. The PS algorithm~\cite{PS73} evaluates $f$ with $O(\sqrt{d})$ non-scalar (ciphertext-ciphertext) multiplications instead of the $d - 1$ multiplications of a direct evaluation. Chen et al.~\cite{CCS19} adapted the algorithm to the Chebyshev basis, which is essential for the numerical stability of large-degree interpolants, by replacing the power recurrences with $T_{2i} = 2T_i^2 - 1$ and $T_{2i+1} = 2 T_i T_{i+1} - T_1$ and by performing the polynomial long divisions directly in the Chebyshev basis using $2 T_i T_j = T_{i+j} + T_{|i-j|}$. Given $k$ and $\mu$ (denoted $m$ in~\cite{PS73,CCS19}; we use $\mu$ because $m$ denotes messages elsewhere in this paper) with $k(2^\mu - 1) > d$, the algorithm proceeds as follows.
\begin{enumerate}
  \item \emph{Baby steps.} Compute the basis $\mathcal{B} = \{B_1(x), \ldots, B_k(x)\}$ with $k - 1$ non-scalar multiplications (by repeated squaring and products of existing elements).
  \item \emph{Giant steps.} Compute $\mathcal{G} = \{B_k(x), B_{2k}(x), \ldots, B_{2^{\mu-1}k}(x)\}$ by $\mu - 1$ squarings, and $B_{k(2^\mu-1)}(x)$ by $\mu - 1$ further products of elements of $\mathcal{G}$.
  \item \emph{Recursion.} Make $f$ monic of degree $k(2^\mu - 1)$ by adding $B_{k(2^\mu-1)}$ (subtracted again at the end), and write it by long division as
  \begin{equation}
    \label{eq:ps}
    f = \left( B_{2^{\mu-1}k} + f_c \right) \cdot f_q + f_s, \qquad \deg f_c < k, \quad \deg f_q = \deg f_s = k(2^{\mu-1} - 1),
  \end{equation}
  where $f_q$ and $f_s$ are monic and are evaluated recursively with $\mu - 1$ in place of $\mu$. The recursion stops when the degree drops to $k$, at which point a polynomial is a \emph{leaf}: a linear combination $\sum_{i \le k} w_i B_i(x) + w_0$ of the baby-step basis with public scalar weights, which costs no non-scalar multiplication. Likewise, $f_c$ is a leaf at every node.
\end{enumerate}
The recursion tree has $2^{\mu-1} - 1$ internal nodes, each costing one non-scalar multiplication, and $2^\mu - 1$ leaves (including the $f_c$ polynomials). For $k \approx \sqrt{d/2}$, the total is $k + 2\mu + 2^{\mu-1} - 4 \approx \sqrt{2d} + \log d$ non-scalar multiplications at multiplicative depth $\lceil \log k \rceil + \mu$, which equals $\lceil \log d \rceil$ or $\lceil \log d \rceil + 1$; Table~\ref{tab:ps-depth} lists the depth for the degrees relevant to this work.

\begin{table}[htbp]
  \centering
  \caption{Multiplicative depth of the PS evaluation of a (non-monic) polynomial of degree $d$ with the best choice of $(k, \mu)$; the depth excludes the affine transformation of the input to $[-1, 1]$ in the Chebyshev case.}
  \label{tab:ps-depth}
  \small
  \begin{tabular}{lccccccccc}
    \toprule
    $d$ & 6--13 & 14--27 & 28--59 & 60--119 & 120--247 & 248--495 & 496--1007 & 1008--2031 & 2032--4031 \\
    \midrule
    depth & 4 & 5 & 6 & 7 & 8 & 9 & 10 & 11 & 12 \\
    \bottomrule
  \end{tabular}
\end{table}

In bootstrapping, the same algorithm is used in two settings. \textsc{EvalExp} (and \textsc{EvalMod} in regular bootstrapping) evaluates a real Chebyshev interpolant over an interval $[-K, K]$, which requires an affine transformation of the input to $[-1, 1]$; this transformation costs a level unless the interval is symmetric, in which case it is a constant multiplication folded into the preceding operation, or of width two, in which case it is a free shift. \textsc{EvalLUT} and \textsc{EvalSeries} evaluate a power series $\sum_{j=1}^{d} c_j z^j$ in the root of unity $z$ with complex coefficients $c_j$ (the trigonometric coefficients of Section~\ref{sec:prelim}), of which only the real part is needed; here the baby and giant steps are plain powers of $z$, and the leaves are weighted sums with complex weights, which CKKS supports natively by multiplying with complex plaintext constants.

\subsection{Algorithmic Improvements}
\label{sec:ps-impr}

In the reference OpenFHE implementation for the \textsc{FlexibleAuto} scaling technique (Algorithm~\ref{alg:ps-ref}), every leaf and every internal node of the PS recursion is treated as an independent CKKS computation. The baby-step basis is first brought to the level and scaling factor of $B_k$ by level adjustments (a constant multiplication followed by a rescaling) and is kept at noise degree two. Every weighted sum then rescales a fresh copy of each of its operands before multiplying it by its weight and leaves the sum at noise degree two, and a further adjustment combines the sum with the giant-step element $B_{2^{\mu-1}k}$ (for $f_c$) or with a copy of $B_k$ that carries the leading coefficient (for the monic $f_q$ and $f_s$). At an internal node, both operands of the product in~\eqref{eq:ps} are rescaled on copies, and the value of $f_s$ is adjusted to the level and scaling factor of the product. Each of these adjustments costs a scalar multiplication, a ciphertext copy, and usually a rescaling, and each rescaling introduces a rounding error that is amplified by all multiplications above it in the tree. Our algorithm removes almost all of them.

The rest of this section describes our optimizations: weighted sums that are produced directly at a target level and scaling factor, a placement of rescalings and relinearizations that avoids redundant operations (Algorithm~\ref{alg:ps-ours}), and a batched evaluation of the leaves.

\paragraph{Weighted sums with a target level and scale.}
The elements of $\mathcal{B}$ live at different levels, and a leaf is evaluated $2^\mu - 1$ times with different weights, target levels, and target scaling factors. Instead of adjusting the operands, we adjust the weights: a weighted sum is produced directly at a chosen level $\ell$ and scaling factor $\sigma$ by encoding, for a term $w_i B_i(x)$ with $B_i(x)$ carrying scaling factor $\Delta(B_i)$, the constant $w_i \cdot \sigma / \Delta(B_i)$ on the fly and dropping the towers of $B_i$ beyond $Q_\ell$, which is exact. Every term of the sum then lands at level $\ell$ with scaling factor $\sigma$, and the constant term $w_0$ is added at that scale. A sum that is about to be multiplied is produced one level above its consumer, with $\sigma$ of the form $q_{\ell} \Delta$, so that a single rescaling yields the operand at the level below with scaling factor $\Delta$; a sum on an $f_s$ branch is instead produced directly at the level and the noise-degree-two scaling factor of the product it is added to, and is never rescaled on its own. The target level and scaling factor of each node are tracked through the recursion, the $f_q$ branch targeting one level above its parent and the $f_s$ branch inheriting the target of its parent, so that the two operands of every product and both summands of every addition in the composition tree meet at the same level and scale without any adjustment.

\begin{algorithm}[!htbp]
  \caption{Reference evaluation of one PS node (OpenFHE v1.6.1, \textsc{FlexibleAuto}). $\textsc{Eval}$ dispatches to $\textsc{WSum}$ for a leaf ($\deg f \le k$) and to $\textsc{Node}$ otherwise. All elements of $\mathcal{B}$ have been adjusted to the level of $B_k$ and left at noise degree two; every non-scalar multiplication, in the baby and giant steps as well, is relinearized as soon as it is formed.}
  \label{alg:ps-ref}
  \begin{algorithmic}[1]
    \Function{WSum}{$w$}
      \State $\ell^\ast \gets$ level of the deepest $\bar{B}_i$ with $w_i \ne 0$
      \For{each $i$ with $w_i \ne 0$}
        \State $t \gets \mathsf{Adjust}(\mathsf{Copy}(\bar{B}_i), \ell^\ast)$; \; $t \gets \mathsf{RS}(t)$; \; $\bar{t} \gets w_i \cdot t$ \Comment{one rescaling per term}
        \State $\bar{v} \gets \bar{v} + \bar{t}$
      \EndFor
      \State \Return $\bar{v} + w_0$ \Comment{noise degree two}
    \EndFunction
    \Function{Eval}{$f, k, \mu$}
      \If{$\deg f \le k$}
        \State \Return \Call{WSum}{$f$} \Comment{leaf: the weights are the coefficients of $f$}
      \Else
        \State \Return \Call{Node}{$f, k, \mu$}
      \EndIf
    \EndFunction
    \Function{Node}{$f, k, \mu$}
      \State $(f_c, f_q, f_s) \gets$ long division of $f$ by $B_{2^{\mu-1}k}$ and of the remainder by $f_q$
      \State $\bar{v}_q \gets$ \Call{Eval}{$f_q, k, \mu-1$}; \; $\bar{v}_s \gets$ \Call{Eval}{$f_s, k, \mu-1$}
      \State $\bar{v}_c \gets$ \Call{WSum}{$f_c$}; \; $\bar{v}_c \gets \mathsf{Adjust}(\bar{v}_c, \mathrm{level}(B_{2^{\mu-1}k})) + B_{2^{\mu-1}k}$
      \State $\bar{v} \gets \mathsf{Relin}\big(\mathsf{RS}(\mathsf{Copy}(\bar{v}_c)) \otimes \mathsf{RS}(\mathsf{Copy}(\bar{v}_q))\big)$ \Comment{operands rescaled on copies; every product relinearized at once}
      \State \Return $\bar{v} + \mathsf{Adjust}(\mathsf{Copy}(\bar{v}_s), \mathrm{level}(\bar{v}))$ \Comment{$\bar{v}_s$ adjusted (rescaled) to the product}
    \EndFunction
  \end{algorithmic}
\end{algorithm}

\paragraph{Rescaling placement.}
We reorder the computation so that a rescaling is performed only where a non-scalar multiplication requires a noise-degree-one operand, so that every rescaling is shared by as many terms as possible, and so that a relinearization is performed only when its result is about to be multiplied again. Algorithms~\ref{alg:ps-ref} and~\ref{alg:ps-ours} contrast the reference and our evaluation of one node of the recursion tree; $\mathsf{RS}$ denotes a rescaling, $\mathsf{Adjust}(\cdot, \ell)$ a level/scale adjustment to level $\ell$, $\otimes$ a non-scalar multiplication without relinearization, $\mathsf{Relin}$ a relinearization, $\mathsf{RelinRS}$ a relinearization fused with a rescaling (Section~\ref{sec:fused}), and a bar marks an unrescaled (noise-degree-two) element.

\begin{algorithm}[!htbp]
  \caption{Our evaluation of one PS node. Levels follow Section~\ref{sec:prelim}: a ciphertext at level $\ell$ has modulus $Q_\ell$, and a rescaling moves it to level $\ell - 1$. $\textsc{WSum}(w, \ell, \sigma)$ produces a weighted sum of basis elements directly at level $\ell$ and scaling factor $\sigma$; $\textsc{Node}(f, k, \mu, \ell, \sigma)$ returns the value of $f$, neither relinearized nor rescaled, at level $\ell$ with (noise-degree-two) scaling factor $\sigma$. $\textsc{Eval}$ dispatches to $\textsc{WSum}$ for a leaf ($\deg f \le k$) and to $\textsc{Node}$ otherwise, and $\textsc{EvalPS}$ is the entry point. $\Delta(\mathbf{ct})$ and $\ell(\mathbf{ct})$ denote the scaling factor and the level of a ciphertext.}
  \label{alg:ps-ours}
  \begin{algorithmic}[1]
    \Function{WSum}{$w, \ell, \sigma$}
      \State $\bar{v} \gets w_0$ encoded at scale $\sigma$ at level $\ell$
      \For{each $i$ with $w_i \ne 0$}
        \State $\bar{v} \gets \bar{v} + \big( w_i \, \sigma / \Delta(B_i) \big) \cdot B_i$ \Comment{towers of $B_i$ beyond $Q_\ell$ dropped; no adjustment, no rescaling}
      \EndFor
      \State \Return $\bar{v}$ \Comment{level $\ell$ and scale $\sigma$, whatever the levels of the $B_i$}
    \EndFunction
    \Function{Eval}{$f, k, \mu, \ell, \sigma$}
      \If{$\deg f \le k$}
        \State \Return \Call{WSum}{$f, \ell, \sigma$} \Comment{leaf: the weights are the coefficients of $f$}
      \Else
        \State \Return \Call{Node}{$f, k, \mu, \ell, \sigma$}
      \EndIf
    \EndFunction
    \Function{Node}{$f, k, \mu, \ell, \sigma$}
      \State $(f_c, f_q, f_s) \gets$ long division of $f$ by $B_{2^{\mu-1}k}$ and of the remainder by $f_q$; \; $\sigma_q \gets \sigma q_{\ell+1}^2 / \Delta(\bar{B}_{2^{\mu-1}k})$
      \State $\bar{v}_q \gets$ \Call{Eval}{$f_q, k, \mu-1, \ell+1, \sigma_q$}
      \If{$\deg f_q \le k$}
        \State $v_q \gets \mathsf{RS}(\bar{v}_q)$ \Comment{a leaf has no quadratic term to relinearize}
      \Else
        \State $v_q \gets \mathsf{RelinRS}(\bar{v}_q)$ \Comment{the only relinearization and rescaling of this branch}
      \EndIf
      \State $\bar{v}_s \gets$ \Call{Eval}{$f_s, k, \mu-1, \ell, \sigma$} \Comment{neither rescaled nor relinearized}
      \State $v_c \gets \mathsf{RS}\big(\textsc{WSum}(f_c, \ell(\bar{B}_{2^{\mu-1}k}), \Delta(\bar{B}_{2^{\mu-1}k})) + \bar{B}_{2^{\mu-1}k}\big)$ \Comment{one rescaling shared with the giant step}
      \State \Return $v_c \otimes v_q + \bar{v}_s$ \Comment{unrelinearized product at $(\ell, \sigma)$, towers of $v_c$ beyond $Q_\ell$ dropped; the quadratic term of $\bar{v}_s$ is accumulated}
    \EndFunction
    \Function{EvalPS}{$f, k, \mu$}
      \State $\bar{v} \gets \textsc{Node}\big(f, k, \mu, \ell(\bar{B}_{k(2^\mu - 1)}), \Delta(\bar{B}_{k(2^\mu - 1)})\big) - \bar{B}_{k(2^\mu - 1)}$ \Comment{the subtracted top element is itself unrelinearized and unrescaled}
      \State \Return $\mathsf{Relin}(\bar{v})$ \Comment{single relinearization at the root, shared with the last giant step; the rescaling is left to the consumer}
    \EndFunction
  \end{algorithmic}
\end{algorithm}

\emph{Baby and giant steps.} The recurrences subtract their constant or $T_1$ term before rescaling, i.e., $T_{2i} = \mathsf{RS}(2T_i^2 - 1)$ and $T_{2i+1} = \mathsf{RS}(2 T_i T_{i+1} - T_1)$. The reference realigns every baby-step element to the level of $B_k$ and rescales a fresh copy of it at every use; we keep the elements at the levels of their own depths and absorb the level differences into the weights of the leaves. We also keep unrescaled copies $\bar{B}_k$ and $\bar{B}_{2^j k}$, $1 \le j < \mu$, of the top baby-step element and of the giant-step elements, and leave the top element $\bar{B}_{k(2^\mu-1)}$ unrescaled and unrelinearized, as it is only subtracted from the root.

\emph{Internal nodes.} The value $v_c$ of $f_c$ is added to the unrescaled $\bar{B}_{2^{\mu-1}k}$, and the sum is rescaled once. The value of $f_q$ is rescaled, since it is about to be multiplied, whereas the value of $f_s$ is added to the product before the product is rescaled. A node thus returns the unrescaled
\begin{equation}
  \label{eq:ps-node}
  \bar{v} = \mathsf{RS}\big( v_c + \bar{B}_{2^{\mu-1}k} \big) \cdot \mathsf{RS}(\bar{v}_q) + \bar{v}_s,
\end{equation}
and its parent either rescales it (on the $f_q$ branch) or adds it as is (on the $f_s$ branch). Consequently, a leaf is rescaled only on the $f_q$ branch, and no value is rescaled twice.

\emph{Relinearization.} The product in~\eqref{eq:ps-node} is also left unrelinearized: its quadratic term is carried through the addition of $\bar{v}_s$ and removed only when the value is multiplied again or reaches the root. Besides the baby- and giant-step products, only the values on the $f_q$ branches and the root are therefore relinearized, and a relinearization that is immediately followed by a rescaling is fused with it. The root is relinearized once, together with $\bar{B}_{k(2^\mu-1)}$, and its rescaling is left to the operation that consumes the result, as is customary under \textsc{FlexibleAuto}.

Table~\ref{tab:ps-ops} summarizes the resulting operation counts. The number of rescalings of the reference grows linearly with the degree, because every weighted sum rescales a fresh copy of each of its operands. Algorithm~\ref{alg:ps-ours} performs $k + 2\mu - 4$ rescalings in the baby and giant steps, two per internal node (for $v_c$ and $v_q$), and one at the root, i.e., $k + 2\mu + 2^\mu - 5$ in total. The reference relinearizes each of its $k + 2\mu + 2^{\mu-1} - 4$ non-scalar multiplications, whereas we relinearize only the $k + 2\mu - 4$ products of the baby and giant steps, the $2^{\mu-2} - 1$ internal nodes on $f_q$ branches, and the root, i.e., $k + 2\mu - 4 + 2^{\mu-2}$ in total. The scalar multiplications are reduced to the non-zero weights of the leaves, the adjustments of the leaves and of the $f_s$ values (and their copies) are eliminated, and the depth is unchanged.

\begin{table}[!htbp]
  \centering
  \caption{Operation counts of one PS evaluation of a polynomial of degree $d$ (with the best $(k, \mu)$ and dense coefficients) under \textsc{FlexibleAuto} in OpenFHE v1.6.1 (Ref.) and in our implementation, for the power basis and, in parentheses, for the Chebyshev basis where the two differ. The counts were obtained by instrumenting the primitive operations of the library. Adjustments are also included in the scalar-multiplication and rescaling counts, a fused relinearization and rescaling counts as one of each, and the final rescaling of the result is excluded for both implementations.}
  \label{tab:ps-ops}
  \small
  \setlength{\tabcolsep}{4.5pt}
  \begin{tabular}{clcccccc}
    \toprule
    $d$ ($k$, $\mu$) & & \begin{tabular}[c]{@{}c@{}}Non-scalar\\ mult.\end{tabular} & \begin{tabular}[c]{@{}c@{}}Relinear-\\ izations\end{tabular} & Rescalings & \begin{tabular}[c]{@{}c@{}}Scalar\\ mult.\end{tabular} & \begin{tabular}[c]{@{}c@{}}Adjust-\\ ments\end{tabular} & \begin{tabular}[c]{@{}c@{}}Ciphertext\\ copies\end{tabular} \\
    \midrule
    \multirow{2}{*}{27 (4, 3)}    & Ref.  & 10  & 10  & 46 (48)     & 33 (36)     & 12 (15)   & 60 (63)     \\
                                  & Ours  & 10  & 8   & 15 (17)     & 28 (31)     & 3 (6)     & 7 (11)      \\
    \midrule
    \multirow{2}{*}{119 (8, 4)}   & Ref.  & 20  & 20  & 161 (163)   & 131 (136)   & 26 (31)   & 188 (193)   \\
                                  & Ours  & 20  & 16  & 33 (36)     & 120 (126)   & 7 (13)    & 15 (22)     \\
    \midrule
    \multirow{2}{*}{239 (16, 4)}  & Ref.  & 28  & 28  & 303 (300)   & 259 (265)   & 34 (40)   & 323 (329)   \\
                                  & Ours  & 28  & 24  & 42 (45)     & 241 (251)   & 8 (18)    & 16 (27)     \\
    \midrule
    \multirow{2}{*}{495 (16, 5)}  & Ref.  & 38  & 38  & 582 (580)   & 519 (526)   & 54 (61)   & 634 (641)   \\
                                  & Ours  & 38  & 30  & 67 (71)     & 496 (507)   & 15 (26)   & 31 (43)     \\
    \midrule
    \multirow{2}{*}{1007 (16, 6)} & Ref.  & 56  & 56  & 1137 (1136) & 1039 (1047) & 94 (102)  & 1253 (1261) \\
                                  & Ours  & 56  & 40  & 116 (121)   & 1007 (1019) & 30 (42)   & 62 (75)     \\
    \midrule
    \multirow{2}{*}{2031 (16, 7)} & Ref.  & 90  & 90  & 2244 (2244) & 2079 (2088) & 174 (183) & 2488 (2497) \\
                                  & Ours  & 90  & 58  & 213 (219)   & 2030 (2043) & 61 (74)   & 125 (139)   \\
    \midrule
    \multirow{2}{*}{4031 (64, 6)} & Ref.  & 104 & 104 & 4301 (4258) & 4111 (4121) & 142 (152) & 4371 (4381) \\
                                  & Ours  & 104 & 88  & 166 (171)   & 4033 (4069) & 32 (68)   & 64 (101)    \\
    \bottomrule
  \end{tabular}
\end{table}

Since the scaling factor differs from level to level under \textsc{FlexibleAuto}, a few adjustments remain in addition to the operations of Algorithm~\ref{alg:ps-ours}. Each adjustment consists of a scalar multiplication applied to a copy of the ciphertext, usually followed by a rescaling. They occur in three places: (i) in the baby and giant steps, when the two operands of a product sit at different levels (e.g., $B_1$ and $B_2$ in $B_3 = B_1 \cdot B_2$), the shallower one is adjusted to the deeper level; (ii) in the Chebyshev basis, the correction terms $T_1$ and $\bar{T}_k$ of the recurrences are brought to the level of the product they are subtracted from; and (iii) at the internal nodes on $f_s$ branches, the giant-step element cannot share the rescaling of $v_c$ and is rescaled on a copy. None of these adjustments affect a leaf or an $f_s$ value. For a degree-$119$ polynomial ($k = 8$, $\mu = 4$), the number of rescalings drops from $161$ to $33$ in the power basis and from $163$ to $36$ in the Chebyshev basis, and the number of relinearizations from $20$ to $16$. The output differs from that of the reference only in the rounding order, at essentially the same precision (within half a bit).

\paragraph{Batched evaluation of the leaves.}
The $2^\mu - 1$ leaves depend only on $\mathcal{B}$, on public weights, and on target levels, all of which are determined by the long divisions, which are pure scalar computations. We therefore evaluate the recursion in two passes. A recording pass performs the long divisions and collects, for every leaf, its weight vector (with the bias in position $0$ and, where applicable, the giant-step element as an extra weight) and its target level, without touching any ciphertext. All recorded leaves are then evaluated at once by a single batched weighted-sum operation, which reads each element of $\mathcal{B}$ once and accumulates its contribution to every leaf, for real and for complex weights alike. A replay pass finally walks the composition tree, consuming the precomputed leaves in the recorded order and performing only the $2^{\mu-1} - 1$ non-scalar multiplications of the internal nodes and the $2^{\mu-2}$ relinearizations of the $f_q$ branches and the root. The batched pass turns $2^\mu - 1$ memory-bound passes over the baby-step basis into one, and removes the corresponding kernel launches on the GPU; the replay pass contains the entire sequential dependency chain of the evaluation and little else: the leaf rescalings are performed all at once at its start, and the internal nodes of one height of the tree are independent of each other, so the replay proceeds in $\mu - 1$ waves of parallel node evaluations, just as the baby steps of one doubling round are computed in parallel. The same two-pass structure is used for the Chebyshev and the power basis, so that \textsc{EvalExp}, \textsc{EvalLUT}, and \textsc{EvalSeries} share one evaluator.

The improvements of this section are implemented in the OpenFHE CPU code path as well, where they make the same decisions, operation by operation, as our GPU implementation (Section~\ref{sec:gpu}), so that the two produce bit-identical ciphertexts; Section~\ref{sec:results-ps} quantifies their effect on the latency of the polynomial evaluation.

\FloatBarrier
\section{Algorithmic Bootstrapping Optimizations}
\label{sec:algopt}

This section describes the CKKS bootstrapping optimizations that we developed on top of OpenFHE v1.6.1~\cite{OpenFHE161}. All of them apply to both our CPU (OpenFHE) and GPU (FIDESlib) implementations. Some of these optimizations apply to all variants of CKKS bootstrapping, whereas others are specific to a particular flavor, e.g., continuous functional bootstrapping. Likewise, some of them address specifics of the OpenFHE bootstrapping pipeline, which FIDESlib follows, whereas others apply to any CKKS bootstrapping pipeline; we provide clarifications where needed. The common theme of most of these optimizations is to reduce the multiplicative depth of functional bootstrapping, so that smaller parameters can be used.

\subsection{Saving a Level at Modulus Raising}
\label{sec:omr}

\paragraph{OpenFHE setting.}
We use hybrid key switching~\cite{HK20} (see Appendix~B of the full version of~\cite{KPZ21} for a detailed description) with special modulus $P$, a product of special primes $p_i$, and decomposition number $\beta$: all evaluation keys are RLWE samples over $Q_L P$, and this modulus determines the security level. A key switching of a ciphertext at modulus $Q_\ell$ decomposes its second component into $\beta$ digits, multiplies them by the $\beta$ rows of the key over $Q_\ell P$, and divides the result by $P$ with rounding, which adds a noise term $\mathbf{e}_{\mathrm{ks}}$ whose magnitude does not depend on the current modulus. Throughout this section, the division of a ciphertext by a modulus, whether in a rescaling or at the end of a key switching, denotes a division with rounding to the nearest integer, applied coefficient-wise. After \textsc{ModRaise}, the ciphertext decrypts to $\mathbf{t} = \mathbf{m} + q_0 \mathbf{I}$ and is multiplied by a public normalization constant $c \in (0, 1)$ before \textsc{CtS}, so that the slots entering \textsc{EvalMod}, \textsc{EvalExp}, or \textsc{EvalSeries} are of the form $\Delta \cdot (m_i/q_0 + I_i)$. This separate normalization step is specific to the OpenFHE implementation of CKKS bootstrapping, where $c = 2^{-\eta}/N$ with $2^\eta = q_0/\Delta$ (the factor $1/N$ accounts for the unnormalized encoding matrices of \textsc{CtS}), and $c = 1/N$ in the discrete case, where $\Delta = q_0$. Like any other constant multiplication, it requires a rescaling by $q_L$ and hence consumes one level before \textsc{CtS} even starts. For sparse packing ($n < N/2$), the $N/(2n)$ copies of the lifted message also have to be combined by the partial sum (trace)
\begin{equation}
  \label{eq:partialsum}
  \mathbf{ct} \leftarrow \sum_{i=0}^{N/(2n)-1} \mathrm{Rot}(\mathbf{ct}, i \cdot n)
\end{equation}
before \textsc{CtS}. It is implemented as a radix-$r$ fold of hoisted rotations: each of the $\log_r(N/(2n))$ rounds computes the digit decomposition once and reuses it for $r - 1$ rotations, whose key-switched products are summed before a single modulus reduction. A radix larger than two (we use $r = 4$) amortizes the decomposition and the modulus reduction over more rotations and reduces the number of sequential rounds, at the cost of more rotation keys.

\paragraph{OverModRaise for the normalization constant.}
The OverModRaise approach~\cite{KCY25} observes that \textsc{ModRaise} can lift the ciphertext to a modulus $Q_{\mathrm{over}} > Q_L$, as long as every key-dependent operation is performed at a modulus not exceeding $Q_L P$. Two ways of consuming the extra modulus are proposed in~\cite{KCY25}: OverModRaise1 spends it on the first linear transformation of \textsc{CtS}, evaluated multiplication-first at the cost of forgoing the baby-step giant-step algorithm in that layer; OverModRaise2 rescales the lifted ciphertext by the extra modulus without any multiplication, which requires a slightly larger scaling factor in \textsc{CtS} and an inverse rescaling afterwards. Our variant consumes the extra modulus with the normalization constant: like OverModRaise2, it rescales immediately after the lift, but since the rescaling is paired with a multiplication that the pipeline performs anyway, the message leaves the step at its intended scale and \textsc{CtS} is unchanged. Let $A = p_0 \cdots p_{a-1}$ be the product of the first $a$ special primes, chosen so that $\log A \ge 58$ (one 60-bit special prime, or two 30-bit ones), and let $P' = P/A$. \textsc{ModRaise} lifts the ciphertext $\mathbf{ct}$ to $Q_L A$, the normalization is applied as the integer $C = \lfloor c \cdot A \rceil$, and the result is divided by $A$, i.e., we compute $\lfloor C \cdot \mathbf{ct} / A \rceil \bmod Q_L$, which returns a ciphertext over $Q_L$ at scaling factor $\Delta$ and noise degree one. The output is the same ciphertext as before, one level higher: \textsc{CtS} now starts at level $L$ instead of $L-1$. No new primes are needed, as $Q_L A$ is a sub-basis of $Q_L P$. The mechanism applies to regular, discrete, and continuous bootstrapping alike, and the multiplication-first layer of~\cite{KCY25} can be stacked on top of it to save a further level. Without encapsulation, this is the whole optimization: for full packing, no key switching occurs between \textsc{ModRaise} and \textsc{CtS}, and for sparse packing, the partial sum~\eqref{eq:partialsum} is evaluated after the division by $A$, where its rounding noise remains far below the other noise terms of the lifted ciphertext.

\paragraph{The problem with sparse-secret encapsulation.}
With encapsulation~\cite{BTPH22}, the ciphertext is switched from the main secret $\mathbf{s}$ to a sparse secret $\mathbf{s}'$ of Hamming weight $h$ (e.g., $h = 32$) at modulus $q_0$, lifted under $\mathbf{s}'$ (so that $\|\mathbf{I}\| \le h$), and switched back to $\mathbf{s}$ after the lift. In the standard pipeline, the switch-back is a hybrid key switching at modulus $Q_L$ that takes place \emph{before} the rescaling of the normalization, which attenuates its rounding noise $\mathbf{e}_{\mathrm{ks}}$ together with the message. OverModRaise breaks this arrangement in two ways. A switch-back at modulus $Q_L A$ would require a key over $Q_L A P$, which exceeds the secure modulus. A switch-back after the division by $A$, which is the placement suggested in~\cite{KCY25}, is secure, but its noise $\mathbf{e}_{\mathrm{ks}}$ is then added at full strength to a message that has just been shrunk by the factor $c$: the relative noise grows by $1/c = N 2^{\eta}$, i.e., several bits of precision are lost. The same problem arises for the partial sum under encapsulation.

\paragraph{Switching back over the raised modulus.}
Our solution is to perform the switch-back at modulus $Q_L A$ with a key whose gadget modulus is $P'$ rather than $P$. Let $(\mathbf{c}_0, \mathbf{c}_1) \in R_{Q_L A}^2$ be the lifted ciphertext under $\mathbf{s}'$; since $\mathbf{c}_1$ is the exact lift of an element of $R_{q_0}$, we have $\|\mathbf{c}_1\| < q_0$. We generate the key
\begin{equation}
  \label{eq:raisekey}
  \mathsf{ksk}_{\mathbf{s}' \to \mathbf{s}} = (\mathbf{k}_0, \mathbf{k}_1) \in R_{Q_L P}^2, \qquad \mathbf{k}_0 = -\mathbf{k}_1 \cdot \mathbf{s} + \mathbf{e} + P' \cdot \mathbf{s}',
\end{equation}
i.e., a hybrid key over the usual modulus $Q_L P = Q_L A P'$, but with gadget constant $P'$. The noise introduced by a key switching is dominated by the term $\mathbf{d} \cdot \mathbf{e}/P'$, where $\mathbf{d}$ is a digit of the decomposed ciphertext component, so each digit must be bounded by a small multiple of the gadget modulus; we allow digits of size at most $2^4 P'$. Hence $\mathbf{c}_1$ itself serves as a single digit whenever $q_0 \le 2^4 P'$, with no basis extension, which is always the case for 60-bit special primes and $\beta \ge 2$; otherwise, $\mathbf{c}_1$ is split into several digits, each with its own key row. The switch-back computes
\begin{equation}
  \label{eq:raiseks}
  (\mathbf{c}'_0, \mathbf{c}'_1) = \left( \mathbf{c}_0 + \left\lfloor \tfrac{1}{P'} \mathbf{c}_1 \cdot \mathbf{k}_0 \right\rceil,\; \left\lfloor \tfrac{1}{P'} \mathbf{c}_1 \cdot \mathbf{k}_1 \right\rceil \right) \bmod Q_L A,
\end{equation}
where the products are formed over $Q_L P$ and the division by $P'$ maps $Q_L P$ back to $Q_L A$. Since $\mathbf{c}'_0 + \mathbf{c}'_1 \cdot \mathbf{s} = \mathbf{c}_0 + \mathbf{c}_1 \cdot \mathbf{s}' + \mathbf{c}_1 \mathbf{e}/P' + \mathbf{e}_{\mathrm{ks}} \pmod{Q_L A}$, the result encrypts $\mathbf{t}$ under $\mathbf{s}$ over $Q_L A$ with the usual key-switching noise. Every operation is performed at a modulus not exceeding $Q_L P$, so security is unaffected. The normalization by $C$ and the division by $A$ follow, so that $\mathbf{e}_{\mathrm{ks}}$ is attenuated exactly as in the standard pipeline. The switch-back is also cheaper than a standard key switching, as its key has a single row and no basis extension is needed. If $P' = 1$ (i.e., $A = P$), the optimization cannot be applied; this case, a single special prime with $\beta = 1$, does not arise in practical CKKS bootstrapping configurations.

\paragraph{Fusing the partial sum into the switch-back.}
For sparse packing under encapsulation, we exploit that a key switching can switch and rotate at once, as a rotation key does. Let $\psi_j$ denote the automorphism of $R$ implementing $\mathrm{Rot}(\cdot, j)$, and let $T = \min(N/(2n), r^2)$. For $i = 1, \ldots, T-1$, we generate the keys $\mathsf{ksk}_i = (\mathbf{k}_{i,0}, \mathbf{k}_{i,1})$ as in~\eqref{eq:raisekey} but with target secret $\psi_{in}^{-1}(\mathbf{s})$, and set $\mathsf{ksk}_0 = \mathsf{ksk}_{\mathbf{s}' \to \mathbf{s}}$. The fused switch-back then computes
\begin{equation}
  \label{eq:fused}
  \left( \sum_{i=0}^{T-1} \psi_{in}(\mathbf{c}_0) + \left\lfloor \tfrac{1}{P'} \sum_{i=0}^{T-1} \psi_{in}(\mathbf{c}_1 \cdot \mathbf{k}_{i,0}) \right\rceil,\; \left\lfloor \tfrac{1}{P'} \sum_{i=0}^{T-1} \psi_{in}(\mathbf{c}_1 \cdot \mathbf{k}_{i,1}) \right\rceil \right) \bmod Q_L A,
\end{equation}
which equals $\sum_{i=0}^{T-1} \mathrm{Rot}(\mathbf{ct}, i n)$ under $\mathbf{s}$ over $Q_L A$. The digit decomposition of $\mathbf{c}_1$ is shared by all $T$ key multiplications and only one division by $P'$ is performed, so the first $T$ terms of~\eqref{eq:partialsum} cost $T$ key multiplications and one modulus reduction, which is less than the $\log_r T$ rounds of the hoisted fold they replace; all key-switching noise is again introduced before the division by $A$. The additional keys are $T - 1$ single-row keys ($T \le 16$ for $r = 4$), each about $1/\beta$ of the size of a standard rotation key. If $N/(2n) > r^2$ (very sparse packing), the remaining terms are accumulated by the hoisted fold with stride $Tn$ after the division by $A$. The fused keys are an instance of the aggregated key switching of Yan et al.~\cite{YZC+26}, who merge, in the full-packing encapsulated setting, the switch-back from $\mathbf{s}'$ to $\mathbf{s}$ with the first rotation of \textsc{CtS} into a single key. We apply the same switch-and-rotate principle to the partial sum, i.e., to the sparsely packed case that~\cite{YZC+26} explicitly leaves open, and perform it over the raised modulus $Q_L A$. Note that the partial sum cannot be moved before \textsc{ModRaise}, as the trace has to act on the lifted overflow $q_0 \mathbf{I}$.

\paragraph{Generality.}
The construction above is formulated for the normalization constant because it is the first multiplication after the lift in our pipeline, but it does not depend on that choice: whatever the first multiplication is, a constant or the first \textsc{CtS} layer as in the multiplication-first variant of~\cite{KCY25}, the switch-back~\eqref{eq:raiseks} and the fused partial sum~\eqref{eq:fused} are performed over $Q_L A$ before it. The encapsulated setting, for which~\cite{KCY25} only suggests a switch-back after the rescaling, and the sparsely packed setting, which~\cite{KCY25} does not consider, are thus handled without any loss of precision or security, while retaining the one-level saving. The level-conserving rescaling of Yan et al.~\cite{YZC+26} saves a level at the same point of the pipeline by a different mechanism: since the coefficients of the lifted ciphertext are small, the first multiplication of \textsc{CtS} can be rescaled by $q_L$ without dropping that prime from the modulus. It also requires that no key switching precedes the rescaling, which is why~\cite{YZC+26} is restricted to full packing; our technique removes this restriction for the OverModRaise variant, and the two techniques can be combined.

\subsection{Saving Another Level in Continuous Functional Bootstrapping}
\label{sec:stc-fold}

The second optimization is specific to the bootstrapping variants that start with \textsc{StC}, i.e., continuous functional bootstrapping and the \textsc{StC}-first variant of regular bootstrapping. In the OpenFHE implementation, \textsc{ModRaise} is preceded by an adjustment of the input ciphertext: the message is multiplied by a correction factor $2^{-\gamma}$ (which emulates a larger gap between $q_0$ and the message; $\gamma = 0$ in continuous functional bootstrapping) and by the ratio between the scaling factor expected at the raised level and the current one. This adjustment is a constant multiplication followed by a rescaling and therefore consumes one level; moreover, this level is not accounted for in the depth of the bootstrapping procedure, so that one multiplicative level less than expected is available between two consecutive bootstrapping invocations. In the \textsc{StC}-first variants, however, the last operation before \textsc{ModRaise} is already a plaintext-ciphertext multiplication by the last \textsc{StC} matrix. Since both factors are public constants fixed at setup time, we fold their product into this matrix and let \textsc{StC} end directly on the base modulus $q_0$. The separate adjustment and its rescaling are thus eliminated, and a depth budget of $L_{\mathrm{boot}} + L_0$, where $L_{\mathrm{boot}}$ is the depth of the bootstrapping procedure, now yields exactly $L_0$ multiplicative levels between two bootstrapping invocations. The output precision is unaffected, as the same multiplication is performed with the same rounding, only at a different place. Discrete functional bootstrapping does not benefit from this optimization, as its input is an RLWE ciphertext that is lifted directly from $q_0$.

\subsection{Saving Up to Two More Levels in Continuous Functional Bootstrapping}
\label{sec:dfefbt}

The third optimization further reduces the depth of continuous functional bootstrapping when the scaling factor is small, i.e., when the output precision is limited by the noise rather than by the quality of the approximation. Two coarser approximations then become admissible, each saving one level. First, in \textsc{EvalExp}, we approximate the complex exponential of twice the angle by a Chebyshev interpolant of a slightly higher degree and apply one double-angle squaring fewer. The two degrees lie in the same depth interval of Table~\ref{tab:ps-depth}, so the new interpolant costs no additional depth, and one squaring level is saved. The truncation error of the shorter iteration is far below the noise at small scaling factors but would be visible at larger ones, so this variant is enabled only for small scaling factors. Second, at such precisions, a degree-3 Fourier extension already reaches the noise level, and a degree-3 series $c_3 z^3 + c_2 z^2 + c_1 z + c_0$ can be evaluated in two levels instead of three as $z^2 \cdot (c_3 z + c_2) + (c_1 z + c_0)$. Together, the two changes reduce the depth of continuous functional bootstrapping at small scaling factors by up to two levels; each saved level removes one prime from the modulus chain, which can be spent on smaller parameters or on additional levels between bootstrapping invocations.

\subsection{Increasing the Ciphertext Modulus Bound with Uniform Ternary Secrets}
\label{sec:uniform}

The fourth optimization concerns the choice of the main secret in sparse-secret encapsulation. The encapsulated bootstrapping of OpenFHE v1.6.1 (following~\cite{BTPH22}) uses a sparse ternary main secret of Hamming weight $h = 192$ and switches to a sparser secret of weight $32$ only for the modulus raising. The main secret determines the maximum modulus $Q_L P$ admissible for a given ring dimension and security level, and this bound is substantially lower for sparse secrets than for uniform ternary ones: for $N = 2^{16}$ and 128-bit security, $\log(Q_L P)$ is $1553$ bits for $h = 192$ (the bound used in~\cite{AKP25}) and $1747$ bits for uniform ternary secrets~\cite{BCC+25}. We therefore introduce an encapsulation mode in which the main computation is performed under a uniform ternary secret, while the sparse secret of weight $32$ is used, exactly as before, only between the two key switchings around \textsc{ModRaise} (including the fused switch-back of Section~\ref{sec:omr}). The additional $194$ bits of modulus translate directly into additional levels between two bootstrapping invocations or into a smaller ring dimension for the same depth. Two remarks are in order. First, the noise added by rescaling is proportional to the norm of the secret, so it is larger under a uniform ternary secret; for small scaling factors, part of the gained modulus has to be spent on larger scaling factors, which reduces the benefit of this mode. Second, this mode improves the security assurance of the parameters: the uniform ternary tables of~\cite{BCC+25} are considerably more stable than the sparse-secret estimates, which have been repeatedly revised as new attacks on sparse secrets appeared, and the sparse secret used for encapsulation can be chosen with a work factor well above $128$ bits (e.g., $145$) at almost no cost in efficiency, since it only affects the modulus-raising steps.

\subsection{Fused Key Switching and Rescaling}
\label{sec:fused}

In contrast to the previous optimizations, the last optimization of this section does not reduce the depth but rather the number of number-theoretic transforms (NTTs), which dominate the cost of key switching. A hybrid key switching of a ciphertext at modulus $Q_\ell$ ends with an approximate modulus reduction of the key-switched product from the extended basis $Q_\ell P$ back to $Q_\ell$: the $P$-part of the product is converted to coefficient representation, extended to the basis of $Q_\ell$, transformed back to the evaluation representation for every prime of $Q_\ell$, and subtracted from the $Q_\ell$-part before multiplying by $P^{-1}$ (we write $x_i \mapsto (x_i - X'_i) \cdot P^{-1} \bmod q_i$ for the $i$-th residue). In CKKS, the result is very often rescaled immediately afterwards, which repeats the same pattern for the dropped prime $q_\ell$: its residue is converted to coefficient representation, lifted to every remaining prime, transformed back, subtracted, and multiplied by $q_\ell^{-1}$. Each of the two steps thus applies an NTT to every remaining residue. Since both steps are $\mathbb{Z}_{q_i}$-linear and the NTT is a ring isomorphism, the combined map
\begin{equation*}
  x_i \;\mapsto\; \left( (x_i - X'_i) \cdot P^{-1} - [d]_{q_i} \right) \cdot q_\ell^{-1} \;=\; \left( x_i - \mathrm{NTT}\!\left(X'_i + P \cdot [d]_{q_i}\right) \right) \cdot (P q_\ell)^{-1} \pmod{q_i},
\end{equation*}
where $[d]_{q_i}$ is the lift of the dropped residue, can be assembled in coefficient representation and transformed only once per remaining residue. The fused operation performs $|P| + |Q_\ell|$ NTTs per polynomial instead of $|P| + 2|Q_\ell| - 1$, with the same number of modular multiplications and additions, and its output is bit-identical to that of the unfused sequence; in particular, no additional rounding noise is introduced, in contrast to treating $P q_\ell$ as a single basis to switch down from. The fusion is exposed in two forms: a relinearization (or rotation) fused with the rescaling of its result and, for hoisted accumulations, the single modulus reduction that settles a sum accumulated in the extended basis $Q_\ell P$ fused with the rescaling of that sum. We apply it wherever a key switching is immediately followed by a rescaling in the bootstrapping pipeline. In particular, it is used in the linear transformations of \textsc{CtS} and \textsc{StC}, including the final rescaling of their output, and in the polynomial evaluation of \textsc{EvalExp}, \textsc{EvalLUT}, and \textsc{EvalSeries}, for every relinearization that is immediately followed by a rescaling (Section~\ref{sec:ps-impr}).

\section{Efficient GPU Design}
\label{sec:gpu}

Our GPU implementation is based on FIDESlib 2.2.0~\cite{AVG+25}, whose bootstrapping algorithms correspond to those of OpenFHE v1.6.1. We ported all algorithms of Sections~\ref{sec:ps} and~\ref{sec:algopt} to FIDESlib so that the GPU and CPU implementations make the same decisions, operation by operation, and produce bit-identical ciphertexts. This section describes the GPU-specific optimizations. The use of 32-bit limbs (Section~\ref{sec:gpu-32bit}) and the fused kernels for OverModRaise (Section~\ref{sec:gpu-omr}) are new, and the capture and replay of the whole bootstrapping procedure as a CUDA graph (Section~\ref{sec:gpu-graph}) is new in FIDESlib. The sparse plaintext compression (Section~\ref{sec:gpu-compression}) was introduced in Cerium~\cite{JKS+26} and is already implemented in FIDESlib 2.2.0; we summarize it for completeness, as it substantially reduces the memory footprint and the latency of the linear transformations.

\subsection{32-Bit Limbs}
\label{sec:gpu-32bit}

CKKS bootstrapping has traditionally required scaling factors of more than 40 bits for practical precision. Under the \textsc{FlexibleAuto} scaling technique of OpenFHE, the primes $q_i$ of the modulus chain are close to the scaling factor $\Delta$ and therefore typically exceed 32 bits, so OpenFHE and FIDESlib store each RNS limb in a 64-bit word. GPUs, however, have no native 64-bit integer multiplication, so 64-bit modular arithmetic is emulated with 32-bit multiply, multiply-high, and add-with-carry instructions. Functional bootstrapping changes this picture: for small plaintext sizes and for the low-latency continuous setting, a scaling factor of 24 to 30 bits is sufficient (Section~\ref{sec:results}), and all limbs can then be stored in 32-bit words.

We added 32-bit limbs to FIDESlib, together with all the required primitives. A 32-bit limb is used for every prime $p < 2^{30}$. For such primes, the product of two residues is below $2^{60}$ and is computed exactly by a single wide multiplication, and both Shoup's modular multiplication~\cite{Har14} and Barrett reduction~\cite{DQ00} produce partially reduced values in $[0, 4p)$. Moreover, up to 16 products of canonical residues can be accumulated in a 64-bit integer before modular reduction. When the scaling factor is small, we also use 30-bit special primes for hybrid key switching. They are faster than 60-bit special primes in this case, because all limbs of $Q$ and $P$ are then 32-bit. In general, each prime of a context carries its own word size, so a context can mix widths: for example, the primes of $Q$ can be 32-bit while 60-bit special primes are 64-bit. Kernels that operate on a mixed basis process each run of limbs of the same width separately, and the base conversions support all four combinations of input and output widths.

32-bit limbs reduce both the memory footprint and the latency. Every limb of a ciphertext, plaintext, or evaluation key takes 4 bytes per coefficient instead of 8. For a 29-bit bootstrapping configuration at $N = 2^{16}$ with 30 primes in $Q$ and eight 30-bit special primes, the evaluation keys and the (uncompressed) plaintexts of the linear transformations take 6.9~GB, whereas 64-bit words would require 12.8~GB. Ciphertexts and plaintexts take half the memory, and so do the evaluation keys when the special primes are also 30-bit; with 60-bit special primes, which remain 64-bit, the keys take slightly more than half. The arithmetic is also much cheaper. One modular multiplication takes 6 native GPU (SASS) instructions with Shoup reduction and 10 with Barrett reduction on 32-bit limbs, compared to 23 and 38 instructions on 64-bit limbs. For the NTT, the best 64-bit implementation executes 2.9x to 5.7x more instructions than the best 32-bit one, and the 32-bit NTT is 1.5x to 1.8x faster for $N = 2^{14}$ to $2^{17}$. This is why configurations with scaling factors of at most 30 bits are particularly fast on the GPU (Section~\ref{sec:results-discrete}).

\subsection{Fused Kernels for OverModRaise}
\label{sec:gpu-omr}

A direct GPU implementation of the fused switch-back~\eqref{eq:fused} would launch separate kernels for each automorphism, product, and addition, materialize $T$ rotated ciphertexts, and perform a modulus reduction (ModDown) for each term. We instead implement~\eqref{eq:fused} with two fused kernels around a single ModDown. Let $\tilde{c}_0$ and $\tilde{c}_1$ denote the two components of the lifted ciphertext in the evaluation (NTT) domain. In this domain, an automorphism is a permutation of the coefficients. Since the switch-back keys are only used here, we apply the automorphism $\psi_{in}$ to the key $\mathsf{ksk}_i$ once at key-loading time, which gives the keys $(\tilde{k}_{i,0}, \tilde{k}_{i,1})$. The $i$-th term then becomes $\tilde{c}_1[\psi_{in}^{-1}(j)] \cdot \tilde{k}_{i,\cdot}[j]$ for every coefficient index $j$, i.e., the ciphertext is read at permuted indices and the keys are read linearly. The first kernel computes, for every limb $q_i$ of $Q_L P$,
\begin{equation*}
  \left(d_0[j],\, d_1[j]\right) = \left(\sum_{i=0}^{T-1} \tilde{c}_1\!\left[\psi_{in}^{-1}(j)\right] \cdot \tilde{k}_{i,0}[j],\; \sum_{i=0}^{T-1} \tilde{c}_1\!\left[\psi_{in}^{-1}(j)\right] \cdot \tilde{k}_{i,1}[j]\right) \bmod q_i,
\end{equation*}
accumulating the $T$ products in registers with lazy reduction. After a single ModDown of $(d_0, d_1)$ by $P'$, the second kernel adds the permuted first component:
\begin{equation*}
  c'_0[j] = \left(d_0^{\downarrow}[j] + \sum_{i=0}^{T-1} \tilde{c}_0\!\left[\psi_{in}^{-1}(j)\right]\right) \bmod q_i, \qquad c'_1 = d_1^{\downarrow}.
\end{equation*}
No rotated ciphertext is materialized, and all $T$ terms share one ModDown. In addition, the lifted ciphertext is copied directly into a dedicated context for the basis $Q_L A$, and the copies between bases of different limb widths are batched into a single kernel. For a configuration with $T = 16$, the fused implementation needs 2 kernels instead of 109 for a naive implementation and, for $N = 2^{16}$, avoids about 1~GB of intermediate memory traffic.

\subsection{Capturing the Bootstrapping Procedure as a CUDA Graph}
\label{sec:gpu-graph}

One functional bootstrapping in FIDESlib launches 2000 to 3300 kernels, whose average execution time is only 3.5 to 4.9~$\mu$s for the $N = 2^{14}$ parameter sets. Launching a kernel and recording its stream dependencies costs a few microseconds of host time, so the host-side cost of more than 7~ms can exceed the total execution time of the kernels. The latency is then limited by kernel submission rather than by the GPU. CUDA graphs address this problem: a static workload is recorded once and uploaded to the GPU, and it can then be launched (replayed) many times with a single API call. Cerium~\cite{JKS+26} also uses CUDA graphs, which its compiler constructs from the program and memory schedules, and reports a 7\% speedup for bootstrapping. In contrast, we capture the eager execution of an existing library, which requires no compiler support, and we optimize the captured graph as described below.

We capture the whole bootstrapping procedure as a CUDA graph. During capture, FIDESlib executes the bootstrapping as usual, while all kernels, memory copies, and buffer allocations are recorded by the CUDA driver together with their dependencies. A replay must not allocate or free GPU memory if the corresponding free or allocation falls outside the captured region, so we replace all dynamic allocations by offsets into a single static memory pool. The pool is planned from the liveness of the recorded buffers: two buffers share memory only if the dependencies rule out any conflicting access. The kernel arguments and the device-side pointer tables are then rebound to the static pool, and the graph is rebuilt without the allocation nodes. Two further optimizations are applied during the rebuild:
\begin{itemize}
  \item \emph{Upload hoisting.} The host uploads small tables, such as pointer tables and constants, before many kernels. Since their contents do not change, the rebuild performs these uploads once into the static pool and removes the corresponding copy nodes. For example, this removes 303 nodes from a bootstrapping graph and reduces its host launch time from about 950~$\mu$s to 27~$\mu$s.
  \item \emph{Dependency pruning.} The eager execution in FIDESlib records stream dependencies conservatively. Dependency tracking adds a scheduling gap of several microseconds on the GPU, even when many dependencies are redundant. Since the graph exposes all edges explicitly, we remove every edge implied by other edges (transitive reduction). This reduces the number of edges by 80\% to 86\% without changing the order of any two dependent kernels.
\end{itemize}A replayed graph produces the same results as the eager execution. On RTX 4090, graph replay reduces the latency of 1-bit discrete functional bootstrapping at $N = 2^{15}$ from 16.5~ms to 7.2~ms (2.3x), and that of 10-bit discrete functional bootstrapping of $N/2$ values at $N = 2^{16}$ from 58.0~ms to 50.6~ms (1.15x). As expected, the benefit is largest for small parameters, where the latency is dominated by kernel submission; for fully packed continuous functional bootstrapping, whose latency is dominated by kernel execution, the difference is within 2\%. All GPU latencies in Section~\ref{sec:results} are measured with graph replay.

\subsection{Sparse Plaintext Compression}
\label{sec:gpu-compression}

The linear transformations \textsc{CtS} and \textsc{StC} multiply the ciphertext by hundreds of public plaintexts, one per diagonal of the encoding matrices. At $N = 2^{16}$, these plaintexts take several gigabytes, and they are read from GPU memory at every rotation. However, a plaintext that encodes $n$ slots places its $2n$ real-valued coefficients at positions $j g$ with stride $g = N/(2n)$. In the bit-reversed NTT representation, the plaintext is then constant on blocks of $g$ consecutive entries, so it is fully described by one representative per block. This observation was introduced in Cerium~\cite{JKS+26}. FIDESlib stores only these $2n$ representatives per limb and reads entry $i$ as the representative $\lfloor i/g \rfloor$. The compression is applied after the standard encoding, and it is exact; a plaintext is compressed only after checking that it is block-constant, and the results are bit-identical to those of the uncompressed representation. For $g \ge 32$, all 32 threads of a warp read the same representative, and the kernel becomes latency-bound rather than memory-bound; the linear transformation kernel therefore prefetches the representatives into shared memory.

Compression also applies to the bootstrapping of fully packed ciphertexts. Since \textsc{CtS} and \textsc{StC} are factored into several levels of sparse matrices, the diagonals of most levels are periodic with a short period. OpenFHE encodes each diagonal with the nominal number of slots, so we pass the minimal period of each diagonal to the encoding instead. Table~\ref{tab:compression} shows the effect for a fully packed configuration at $N = 2^{16}$: the plaintexts of all levels except the first level of \textsc{CtS} and the last level of \textsc{StC} are compressed by a factor of 32 to 1024, and the total footprint drops from 2.93~GB to 1.01~GB. On RTX 4090, the execution time of the linear transformation kernels drops by 39\%, and the latency of the whole bootstrapping decreases by 4.5\%, from 44.7~ms to 42.7~ms. The gains are larger for sparse packing: for $n = 64$, the plaintexts take 2.25~MB instead of 480~MB.

\begin{table}[!htbp]
  \centering
  \small
  \caption{Size of the plaintexts of the linear transformations per level, without and with compression, for a fully packed configuration ($N = 2^{16}$, $n = 2^{15}$, 29-bit scaling factor). The period is the minimal period of the diagonals of the level.}
  \label{tab:compression}
  \begin{tabular}{l rrrr}
    \toprule
    Level & Period & Full [MB] & Compressed [MB] & Ratio \\
    \midrule
    \textsc{CtS} 0 & 32{,}768 & 598.5 & 598.5 & 1 \\
    \textsc{CtS} 1 & 1024  & 582.8 & 18.2  & 32 \\
    \textsc{CtS} 2 & 32       & 567.0 & 0.6    & 1024 \\
    \textsc{StC} 0 & 32       & 409.5 & 0.4    & 1024 \\
    \textsc{StC} 1 & 1024  & 393.8 & 12.3  & 32 \\
    \textsc{StC} 2 & 32{,}768 & 378.0 & 378.0 & 1 \\
    \midrule
    Total & & 2929.6 & 1008.0 & 2.9 \\
    \bottomrule
  \end{tabular}
\end{table}

\section{Experimental Results}
\label{sec:results}

\subsection{Experimental Setup}
\label{sec:results-setup}

\paragraph{Software.}
The CPU experiments were run using our development branch of OpenFHE, which extends OpenFHE v1.6.1~\cite{OpenFHE161} with the optimizations of Sections~\ref{sec:ps} and~\ref{sec:algopt}. The GPU experiments were run using our extension of FIDESlib 2.2.0~\cite{AVG+25}, which implements the same algorithms on the GPU (Section~\ref{sec:gpu}); keys are generated and data is encrypted with OpenFHE on the host, and the functional bootstrapping itself is executed entirely on the GPU. The CPU code was compiled using clang 18 with native optimizations enabled.

\paragraph{Hardware.}
The CPU experiments were run on a two-socket server with two Intel Xeon Platinum 8360Y (Ice Lake) processors, each with 36 cores, and 128 GB of RAM; the multi-threaded results use 36 threads pinned to one socket. The GPU experiments were run on three NVIDIA GPUs that represent different classes of systems: RTX 4090, a consumer GPU with 24 GB of memory (AMD EPYC 7702P host CPU, NVIDIA driver 580.95, CUDA 13.0); RTX PRO 6000 Blackwell Server Edition, a professional server GPU with 96 GB of memory (AMD EPYC 9575F host CPU, NVIDIA driver 595.71.05, CUDA 13.2); and B200, a datacenter GPU with 180 GB of memory (AMD EPYC 9575F host CPU, NVIDIA driver 595.71.05, CUDA 13.2). All GPU experiments use a single GPU.

\paragraph{Implementation details for full-packing scenarios.}
A CKKS ciphertext of ring dimension $N$ has $N/2$ complex slots and can therefore hold up to $N$ real values when both the real and the imaginary parts of the slots are used. In the discrete case, this corresponds to an RLWE ciphertext whose $N$ coefficients all carry data ($n = N$ in our tables), and in the continuous case to complex inputs whose real and imaginary parts are independent real values. Since the functions that we evaluate are real-valued, the two halves of such a fully packed ciphertext are processed separately in the nonlinear part of functional bootstrapping: \textsc{ModRaise} and the linear transformations (\textsc{CtS} and \textsc{StC}) are performed once, on a single ciphertext, whereas after \textsc{CtS} the real and imaginary parts are split into two ciphertexts, each of which goes through \textsc{EvalExp} and the polynomial evaluation with the same coefficients, and the two results are then recombined into one ciphertext. When at most $N/2$ values are present ($n \le N/2$), they occupy only the real parts of the slots, and a single ciphertext goes through the nonlinear part.

This has two consequences for the results reported below. First, the latency for $N$ values is less than twice the latency for $N/2$ values, as the modulus raising and the linear transformations are not duplicated, so the fully packed case gives the highest throughput per value. Second, reducing the number of values below $N/2$ lowers the cost of the linear transformations, which depends on the number of slots, but leaves the cost of the nonlinear part unchanged, as it operates on ciphertexts of the same ring dimension.

\paragraph{Parameters.}
All experiments use the \textsc{FlexibleAuto} scaling technique. We consider two secret-key configurations: \textsc{Hybrid\_Ternary} (HT), where a uniform ternary secret is used for the main computation and a sparse secret is used only around \textsc{ModRaise} (Section~\ref{sec:uniform}), and \textsc{Sparse\_Encapsulated} (SE), the encapsulated configuration of OpenFHE v1.6.1, where the main secret is sparse. All parameter sets target 128 bits of security. For a given ring dimension, the security level determines the maximum admissible size of the largest modulus $Q_L P$, and a parameter set is accepted only if its $\log (Q_L P)$ does not exceed this bound. For HT, we take the bounds from the 128-bit security table for uniform ternary secrets in~\cite{BCC+25}, i.e., 430 bits for $N = 2^{14}$, 868 bits for $N = 2^{15}$, 1747 bits for $N = 2^{16}$, and 3523 bits for $N = 2^{17}$. For SE, we use the same bounds as in~\cite{AKP25}, i.e., 767 bits for $N = 2^{15}$ and 1553 bits for $N = 2^{16}$. The parameters are also chosen so that the probability of failure of CKKS bootstrapping, i.e., the probability that the integer overflow introduced by \textsc{ModRaise} exceeds the range supported by the approximate modular reduction, is below $2^{-128}$, and so that the probability of an incorrect decryption result in discrete-CKKS functional bootstrapping is below $2^{-40}$. The total number of RNS limbs (of $Q$ and $P$ together) is limited to 64 by the FIDESlib implementation, and the special primes of hybrid key switching are either 30 or 60 bits long.

\paragraph{Latency measurements.}
We report the latency of functional bootstrapping of a single ciphertext, i.e., no batching over multiple ciphertexts is used, and the reported latency does not include key generation, encryption, and decryption. On the GPU, we report the best of 20 runs (graph replays, Section~\ref{sec:gpu-graph}) after one warm-up run; on the CPU, we report the fastest call. The GPU latencies vary little between runs: the median of the 20 runs is typically 0.25\% higher than the best run, at most 1.7\% higher in 95\% of the configurations, and at most 6\% higher in all configurations.

\subsection{Polynomial Evaluation}
\label{sec:results-ps}

We first examine the polynomial evaluation algorithm of Section~\ref{sec:ps} in isolation. Table~\ref{tab:ps-main} compares the latency of one PS evaluation in OpenFHE v1.6.1 and in our implementation on the CPU for the seven degrees analyzed in Table~\ref{tab:ps-ops}, which range from 27 to 4031 (multiplicative depths from 5 to 12), and for both bases. The upper part of this range is representative of discrete functional bootstrapping for larger plaintext sizes, where the interpolation polynomial has a degree of $p - 1$. The experiments use $N = 2^{16}$ and a 59-bit scaling factor.

\paragraph{Single-threaded results.}
The speedup of our implementation increases monotonically with the degree, from 2.2x for $d = 27$ to 10.9x for $d = 4031$ for the power basis, and from 1.8x to 6.8x for the Chebyshev basis. This is the behavior predicted by the operation counts of Table~\ref{tab:ps-ops}. Both implementations perform the same number of non-scalar multiplications and nearly the same number of scalar multiplications, and our implementation saves 14\% to 36\% of the relinearizations and most of the adjustments. The main difference is in the rescalings and ciphertext copies, whose number grows linearly with the degree in the reference implementation and only as $O(\sqrt{d})$ in ours: for the power basis, the number of rescalings is reduced from 46 to 15 for $d = 27$ and from 4301 to 166 for $d = 4031$, and the number of ciphertext copies from 4371 to 64 in the latter case. The measured latencies follow the same trend. When the degree increases by a factor of 149, the latency of v1.6.1 increases by a factor of 93 for the power basis, which is also the factor by which its number of rescalings increases, and by a factor of 67 for the Chebyshev basis, whereas the latency of our implementation increases only by a factor of 19. In our implementation, the Chebyshev series is 1\% to 6\% slower than the power series, which is consistent with the few additional adjustments required by the correction terms of the Chebyshev recurrences (Section~\ref{sec:ps-impr}); otherwise, the two bases share one evaluator. In v1.6.1, in contrast, the power series with complex coefficients is 14\% to 57\% slower than the Chebyshev series, which is why the speedup is higher for the power basis.

\paragraph{Multi-threaded results.}
With 36 threads, our implementation is 4.0x to 8.9x faster for the power basis and 3.4x to 5.8x faster for the Chebyshev basis. Here the speedup is not monotonic: for the power basis, it reaches 6.0x for $d = 119$, drops to 4.5x for $d = 239$ and $d = 495$, and grows again to 8.9x for $d = 4031$. The reason is that our implementation scales uniformly with the number of threads (by a factor of 7.8 to 11.0 over the whole range), as all leaves are evaluated by a single batched weighted-sum operation and the internal nodes of each height of the recursion tree are evaluated in parallel (Section~\ref{sec:ps-impr}), whereas the reference implementation scales poorly for small degrees (by a factor of 4.3 to 4.7 for $d \le 119$) and well for large degrees (by a factor of up to 13.6), where its per-term rescalings and copies are spread over many threads. In absolute terms, a degree-4031 polynomial is evaluated in 1.7 s in either basis with 36 threads, compared to 14.9 s (power) and 10.0 s (Chebyshev) in v1.6.1.

\begin{table}[!htbp]
  \centering
  \caption{Latency [s] of one Paterson--Stockmeyer evaluation of a polynomial of degree $d$ at $N = 2^{16}$ with a 59-bit scaling factor under \textsc{FlexibleAuto}, in OpenFHE v1.6.1 and in our implementation, for 1 thread and for 36 threads. The degrees are those of Table~\ref{tab:ps-ops}; the last three are the largest degrees that can be evaluated at depths 10, 11, and 12 (Table~\ref{tab:ps-depth}). The power series has complex coefficients, as evaluated by \textsc{EvalLUT} and \textsc{EvalSeries}, and the Chebyshev series has real coefficients, as evaluated by \textsc{EvalExp}.}
  \label{tab:ps-main}
  \small
  \begin{tabular}{ll rrr rrr}
    \toprule
    & & \multicolumn{3}{c}{1 thread} & \multicolumn{3}{c}{36 threads} \\
    \cmidrule(lr){3-5} \cmidrule(lr){6-8}
    Basis & $d$ ($k, \mu$) & v1.6.1 & Ours & Speedup & v1.6.1 & Ours & Speedup \\
    \midrule
    Power & 27 ($4, 3$) & 2.166 & 0.975 & 2.22x & 0.494 & 0.124 & 3.97x \\
     & 119 ($8, 4$) & 7.284 & 2.248 & 3.24x & 1.556 & 0.261 & 5.95x \\
     & 239 ($16, 4$) & 13.345 & 3.447 & 3.87x & 1.893 & 0.417 & 4.54x \\
     & 495 ($16, 5$) & 26.614 & 5.107 & 5.21x & 2.636 & 0.585 & 4.51x \\
     & 1007 ($16, 6$) & 54.564 & 7.857 & 6.94x & 4.641 & 0.867 & 5.36x \\
     & 2031 ($16, 7$) & 114.708 & 13.583 & 8.44x & 8.419 & 1.316 & 6.40x \\
     & 4031 ($64, 6$) & 200.779 & 18.373 & 10.93x & 14.911 & 1.678 & 8.89x \\
    \addlinespace
    Chebyshev & 27 ($4, 3$) & 1.908 & 1.034 & 1.84x & 0.442 & 0.132 & 3.36x \\
     & 119 ($8, 4$) & 5.726 & 2.353 & 2.43x & 1.282 & 0.274 & 4.68x \\
     & 239 ($16, 4$) & 9.963 & 3.587 & 2.78x & 1.576 & 0.435 & 3.62x \\
     & 495 ($16, 5$) & 18.873 & 5.272 & 3.58x & 2.138 & 0.608 & 3.51x \\
     & 1007 ($16, 6$) & 37.216 & 8.027 & 4.64x & 3.488 & 0.892 & 3.91x \\
     & 2031 ($16, 7$) & 76.194 & 13.682 & 5.57x & 5.783 & 1.346 & 4.30x \\
     & 4031 ($64, 6$) & 127.825 & 18.908 & 6.76x & 9.978 & 1.732 & 5.76x \\
    \bottomrule
  \end{tabular}
\end{table}

\FloatBarrier
\subsection{Discrete-CKKS Functional Bootstrapping}
\label{sec:results-discrete}

Table~\ref{tab:discrete-main} shows the latency of discrete-CKKS functional bootstrapping for plaintext sizes from 1 to 12 bits and for different numbers of slots on three GPUs, together with the multi-threaded CPU latency of our OpenFHE implementation as a reference. Depending on the configuration, the GPU implementation is 33x to 71x faster than the multi-threaded CPU one. All experiments use the AKP method~\cite{AKP25}. As in~\cite{AKP25}, the parameters are chosen so that no multiplicative CKKS levels are available after functional bootstrapping, i.e., all levels of the modulus chain are consumed by the functional bootstrapping itself. We ran every experiment for both secret-key configurations, HT and SE (Section~\ref{sec:results-setup}). Unless stated otherwise, the numbers quoted below are for HT; the two configurations are compared at the end of this section.

\begin{table}[!htbp]
  \centering
  \begin{threeparttable}
  \caption{Latency [ms] of discrete-CKKS functional bootstrapping of a single ciphertext with $n$ slots on a multi-threaded CPU (Intel Xeon Platinum 8360Y, 36 threads on one socket) and on a single GPU, compared with the programmable bootstrapping of one value in TFHE-rs~\cite{Zama26} (multi-bit variant, best result over the parameter sets with a failure probability of $2^{-40}$ or $2^{-64}$) on the same platform; the TFHE-rs results cover plaintext sizes of up to 6 bits on the GPUs and up to 8 bits on the CPU, and TFHE-rs has no parameter set for a single bootstrapping of 10 or 12 bits. The first-order trigonometric Hermite interpolation (AKP) method of~\cite{AKP25} is used for discrete-CKKS functional bootstrapping. HT denotes the \textsc{Hybrid\_Ternary} secret-key configuration (uniform ternary main secret, Section~\ref{sec:uniform}) and SE the \textsc{Sparse\_Encapsulated} configuration (sparse main secret); the SE results are only shown for the 4-bit, 6-bit, and 8-bit cases. The scaling factor $\Delta$ depends on the number of slots, and ``$\log \Delta$'' gives its range (in bits) over the entries of a row. No multiplicative CKKS levels are available after functional bootstrapping, and the scaling technique is set to \textsc{FlexibleAuto}.}
  \label{tab:discrete-main}
  \small
  \renewcommand{\arraystretch}{0.86}
  \begin{tabular}{lcccc rrrrr r}
    \toprule
    & & & & & \multicolumn{5}{c}{Ours, number of slots $n$} & \\
    \cmidrule(lr){6-10}
    Platform & Bits & Secret & $\log N$ & $\log \Delta$ & $1$ & $64$ & $1024$ & $N/2$ & $N$ & TFHE-rs \\
    \midrule
    Xeon 8360Y & 1 & HT & 14 & 24--26 & 111 & 131 & 176 & 222 & 308 & 4.88 \\
     & 2 & HT & 15 & 24--29 & 143\tnote{a} & 154\tnote{a} & 285 & 315 & 432 & 5.23 \\
     & \multirow{2}{*}{4} & HT & 15 & 31--33 & 223 & 262 & 329 & 438 & 593 & \multirow{2}{*}{8.07} \\
     &  & SE & 15 & 25--29 & 232 & 277 & 345 & 450 & 619 &  \\
     & \multirow{2}{*}{6} & HT & 15 & 31--37 & 301 & 349 & 445 & 849 & 1117 & \multirow{2}{*}{36.42} \\
     &  & SE & 16 & 28--34 & 328\tnote{b} & 375\tnote{b} & 551\tnote{b} & 859 & 1256 &  \\
     & \multirow{2}{*}{8} & HT & 16 & 36--42 & 792 & 903 & 1101 & 1150 & 1866 & \multirow{2}{*}{523} \\
     &  & SE & 16 & 32--38 & 786 & 884 & 1041 & 1149 & 1747 &  \\
     & 10 & HT & 16 & 39--46 & 1189 & 1281 & 1492 & 1656 & 2740 & -- \\
     & 12 & HT & 16 & 43--50 & 1831 & 2022 & 2289 & 2792 & 4427 & -- \\
    \midrule
    RTX 4090 & 1 & HT & 14 & 25--26 & 2.05 & 2.30 & 3.25 & 3.97 & 5.24 & 2.16 \\
     & 2 & HT & 15 & 24--29 & 2.30\tnote{a} & 2.71\tnote{a} & 4.80 & 5.83 & 7.40 & 2.32 \\
     & \multirow{2}{*}{4} & HT & 15 & 29--33 & 3.65 & 4.59 & 9.34 & 13.0 & 16.6 & \multirow{2}{*}{4.03} \\
     &  & SE & 15 & 25--29 & 3.59 & 4.45 & 5.97 & 7.99 & 9.98 &  \\
     & \multirow{2}{*}{6} & HT & 15 & 32--37 & 7.91 & 9.49 & 12.3 & 25.3 & 31.4 & \multirow{2}{*}{15.73} \\
     &  & SE & 16 & 28--34 & 4.78\tnote{b} & 6.08\tnote{b} & 15.0\tnote{b} & 25.4 & 35.8 &  \\
     & \multirow{2}{*}{8} & HT & 16 & 36--42 & 19.2 & 22.8 & 29.4 & 34.7 & 51.5 & \multirow{2}{*}{--} \\
     &  & SE & 16 & 32--38 & 18.7 & 21.9 & 27.5 & 33.5 & 49.8 &  \\
     & 10 & HT & 16 & 39--46 & 30.7 & 34.4 & 41.1 & 48.9 & 77.6 & -- \\
     & 12 & HT & 16 & 43--46 & 52.8 & 56.9 & 67.9 & oom & oom & -- \\
    \midrule
    RTX PRO 6000 & 1 & HT & 14 & 25--26 & 2.05 & 2.33 & 3.14 & 3.83 & 5.05 & 2.46 \\
     & 2 & HT & 15 & 24--29 & 2.33\tnote{a} & 2.49\tnote{a} & 4.18 & 4.90 & 6.24 & 2.63 \\
     & \multirow{2}{*}{4} & HT & 15 & 29--33 & 3.16 & 3.94 & 7.45 & 9.83 & 12.7 & \multirow{2}{*}{4.55} \\
     &  & SE & 15 & 25--29 & 3.08 & 3.73 & 5.06 & 6.41 & 8.46 &  \\
     & \multirow{2}{*}{6} & HT & 15 & 32--37 & 6.07 & 7.41 & 9.38 & 18.6 & 23.0 & \multirow{2}{*}{18.67} \\
     &  & SE & 16 & 28--34 & 4.24\tnote{b} & 5.05\tnote{b} & 11.3\tnote{b} & 18.0 & 25.4 &  \\
     & \multirow{2}{*}{8} & HT & 16 & 36--42 & 13.5 & 15.8 & 20.0 & 24.0 & 35.5 & \multirow{2}{*}{--} \\
     &  & SE & 16 & 32--38 & 13.1 & 15.5 & 19.8 & 23.1 & 34.3 &  \\
     & 10 & HT & 16 & 39--46 & 20.4 & 23.5 & 28.6 & 32.9 & 51.7 & -- \\
     & 12 & HT & 16 & 43--50 & 36.1 & 39.2 & 46.7 & 54.7 & 88.7 & -- \\
    \midrule
    B200 & 1 & HT & 14 & 25--26 & 2.38 & 2.70 & 3.64 & 4.32 & 5.81 & 0.93 \\
     & 2 & HT & 15 & 24--29 & 2.70\tnote{a} & 2.77\tnote{a} & 4.44 & 5.25 & 6.92 & 0.92 \\
     & \multirow{2}{*}{4} & HT & 15 & 29--33 & 3.63 & 4.17 & 8.02 & 10.1 & 13.7 & \multirow{2}{*}{1.01} \\
     &  & SE & 15 & 25--29 & 3.54 & 4.02 & 5.35 & 6.65 & 9.07 &  \\
     & \multirow{2}{*}{6} & HT & 15 & 32--37 & 7.11 & 7.88 & 10.1 & 17.7 & 23.0 & \multirow{2}{*}{5.01} \\
     &  & SE & 16 & 28--34 & 4.76\tnote{b} & 5.10\tnote{b} & 11.7\tnote{b} & 17.6 & 26.0 &  \\
     & \multirow{2}{*}{8} & HT & 16 & 36--42 & 14.4 & 15.9 & 19.7 & 23.7 & 36.2 & \multirow{2}{*}{--} \\
     &  & SE & 16 & 34--38 & 14.3 & 15.7 & 19.7 & 22.9 & 35.3 &  \\
     & 10 & HT & 16 & 39--46 & 22.2 & 24.1 & 29.0 & 32.6 & 53.4 & -- \\
     & 12 & HT & 16 & 43--50 & 40.3 & 41.5 & 51.1 & 56.8 & 95.5 & -- \\
    \bottomrule
  \end{tabular}
  \begin{tablenotes}\footnotesize
    \item[a,b] Obtained for $N = 2^{14}$ and $N = 2^{15}$, respectively (a smaller ring dimension than for the other entries of the row); oom: every candidate configuration ran out of GPU memory.
  \end{tablenotes}
  \end{threeparttable}
\end{table}

\paragraph{Parameter selection.}
The parameters were selected separately for each GPU, plaintext size, and number of slots. For each configuration, we ran a search over the ring dimension, the decomposition number $\beta$, the level budgets for \textsc{CtS} and \textsc{StC}, the scaling factor, and the size of the special primes (30 or 60 bits), and kept the parameter set with the smallest latency among those that meet the security and failure-probability requirements of Section~\ref{sec:results-setup}. The smallest ring dimension that is sufficient for a given plaintext size is typically optimal: $2^{14}$ for the Boolean case (and for 2 bits with $n \le 64$), $2^{15}$ for 2 to 6 bits, and $2^{16}$ for 8 to 12 bits. The optimal level budgets range from one or two levels for \textsc{CtS} and one level for \textsc{StC} when $n \le 64$ to five levels for \textsc{CtS} and three levels for \textsc{StC} for (nearly) fully packed ciphertexts, and the optimal decomposition number varies between 2 and 7. The special primes are 30 bits long whenever the scaling factor does not exceed 30 bits; for larger scaling factors, both 30-bit and 60-bit special primes were considered, and the 60-bit ones turned out to be preferable in most cases on RTX 4090, whereas the 30-bit ones are optimal in almost all cases on B200.

\paragraph{Lowest latency.}
The lowest latency is achieved in the Boolean case, where the ring dimension of $2^{14}$ is sufficient: a single bit is bootstrapped in 2.05 ms on RTX 4090 and RTX PRO 6000 (2.38 ms on B200). The Boolean latency stays below 5 ms for up to $N/2 = 8192$ bits on all three GPUs and reaches 5.05 ms on RTX PRO 6000 when all $N = 16{,}384$ bits of a ciphertext are bootstrapped at once. Latencies below 5 ms are also achieved for 2-bit plaintexts with up to 1024 slots on all three GPUs (and up to $N/2$ slots on RTX PRO 6000) and for 4-bit plaintexts with up to 64 slots on all three GPUs. For larger plaintext sizes, the lowest latencies for a single value are 4.24 ms for 6 bits and 13.1 ms for 8 bits (both with SE), 20.4 ms for 10 bits, and 36.1 ms for 12 bits, all on RTX PRO 6000, which is the fastest of the three GPUs in most configurations; B200 only matches it in a few configurations with the larger parameter sets and is slower for the small parameter sets that give the lowest latencies.

\paragraph{Dependence on the number of slots.}
The latency grows slowly with the number of slots. In the Boolean case on RTX 4090, the latency increases from 2.05 ms for a single bit to 2.30 ms for 64 bits, 3.25 ms for 1024 bits, and 5.24 ms for all 16,384 bits. In other words, bootstrapping four orders of magnitude more bits increases the latency by a factor of less than three, and the amortized cost drops to about 0.3 $\mu$s per bit. The same behavior is observed for larger plaintext sizes: for 1, 2, 8, 10, and 12 bits, the latency for $n = N$ is only 2.4x to 2.7x higher than for $n = 1$ on all GPUs (3.2x for 2 bits on RTX 4090). For example, a single 8-bit value is bootstrapped in 13.5 ms and 65,536 8-bit values in 35.5 ms on RTX PRO 6000. The ratio is higher, 3.2x to 4.6x, for 4 and 6 bits, mainly because the single-value latency is particularly low there; for 4 bits, for instance, the scaling factor does not exceed 30 bits for $n \le 64$. The savings for sparsely packed ciphertexts come from the \textsc{CtS} and \textsc{StC} linear transformations, which become cheaper and need smaller level budgets (and hence fewer RNS limbs in the whole computation) as $n$ decreases, whereas the cost of the polynomial evaluation in \textsc{EvalExp} and \textsc{EvalLUT} does not depend on $n$ for $n \le N/2$. Going from $n = N/2$ to $n = N$ increases the latency by 32--34\% in the Boolean case and by 62--68\% for 12 bits, as two ciphertexts have to be processed by the polynomial evaluation in the fully packed case, and this evaluation dominates the runtime for larger plaintext sizes.

\paragraph{Comparison with TFHE-rs.}
The last column of Table~\ref{tab:discrete-main} shows the latency of the programmable bootstrapping of a single value in TFHE-rs~\cite{Zama26}, measured on the same platforms. On RTX PRO 6000, the single-value latency of discrete CKKS is lower than that of TFHE-rs for all four plaintext sizes: by a factor of 1.2 for 1 bit (2.05 ms vs.\ 2.46 ms), 1.1 for 2 bits, 1.4 for 4 bits (3.16 ms vs.\ 4.55 ms), and 3.1 for 6 bits (6.07 ms vs.\ 18.67 ms); with SE, the 6-bit factor increases to 4.4 (4.24 ms). On RTX 4090, discrete CKKS is on par with TFHE-rs for 1 to 4 bits (up to 10\% faster) and 2.0x faster for 6 bits (3.3x with SE). On B200, in contrast, TFHE-rs benefits much more from the datacenter GPU: its latency drops to about 1 ms for up to 4 bits and to 5.01 ms for 6 bits, which is 1.4x to 3.6x lower than the single-value latency of discrete CKKS with HT; for 6 bits, discrete CKKS with SE is on par (4.76 ms). These comparisons are for a single value, whereas discrete CKKS bootstraps many values in one operation at a moderately higher latency. For example, on RTX PRO 6000, all 32,768 6-bit values of a fully packed ciphertext are bootstrapped in 23.0 ms, i.e., in 1.2x the time of one 6-bit programmable bootstrapping in TFHE-rs, and on B200, 64 Boolean values are bootstrapped in 2.70 ms and 16,384 in 5.81 ms, compared to 0.93 ms for a single value in TFHE-rs. On the CPU, the single-value latency of TFHE-rs is 8x to 28x lower than that of discrete CKKS for up to 6 bits (e.g., 4.88 ms vs.\ 111 ms for a Boolean value) and 1.5x lower for 8 bits, so that discrete CKKS is advantageous only when many values are bootstrapped at once.

\paragraph{HT vs.\ SE.}
SE has a significant advantage in two cases, both of which are explained by the size of the scaling factor. Scaling factors of at most 30 bits enable a faster GPU implementation based on 32-bit limbs (Section~\ref{sec:gpu-32bit}). For 4 bits, the scaling factor of SE does not exceed 30 bits for any number of slots (it is 25--29 bits), whereas HT requires 30--33 bits and 64-bit limbs for $n = 1024$, $n = N/2$, and $n = N$; for these three numbers of slots, SE is 1.5x to 1.7x faster than HT (e.g., 8.46 ms vs.\ 12.7 ms for a fully packed ciphertext on RTX PRO 6000), while for $n = 1$ and $n = 64$, where the HT scaling factor is 29 bits, the two configurations are within 6\% of each other. For 6 bits, SE needs only 28--29 bits for $n = 1$ and $n = 64$, whereas HT requires 32--34 bits, and SE is 1.4x to 1.7x faster for these two numbers of slots (e.g., 4.24 ms vs.\ 6.07 ms for a single 6-bit value on RTX PRO 6000); for $n \ge 1024$, both configurations need more than 30 bits, and the two are within 3\% of each other or HT is faster, by up to 22\%. This effect is specific to the GPU implementation: on the CPU, HT is faster than SE for 4 bits and for 6 bits in all five cases. For 8 bits, SE is consistently but only slightly faster on the GPUs (by at most 7\%). For all other plaintext sizes, SE offers at most a 5\% benefit, and HT is often faster than SE: it has the lower latency in 38 of the 58 remaining GPU configurations, by up to 42\%, e.g., in the Boolean case, where HT supports the ring dimension of $2^{14}$ while SE needs $2^{15}$. For these reasons, and because HT uses the more conservative uniform ternary secrets for the main computation, we use HT as the main configuration and only include the 4-bit, 6-bit, and 8-bit SE results in Table~\ref{tab:discrete-main}.

\paragraph{Comparison with OpenFHE v1.6.1 on the CPU.}
The CPU results can be compared with the latencies published for OpenFHE v1.6.1~\cite{OpenFHEPerf}, which were measured on the same machine for fully packed ciphertexts ($n = N$) with \textsc{FlexibleAuto} and the SE configuration: 1.4 s, 1.8 s, 2.3 s, and 8.9 s for 1, 2, 4, and 8 bits, respectively (the best multi-threaded results). For the same SE configuration, our implementation takes 426 ms, 431 ms, 619 ms, and 1.75 s, i.e., it is 3.3x, 4.2x, 3.7x, and 5.1x faster. With HT, the corresponding latencies in Table~\ref{tab:discrete-main} are 308 ms, 432 ms, 593 ms, and 1.87 s, which amounts to speedups of 4.5x, 4.2x, 3.9x, and 4.8x over v1.6.1; note that in the Boolean case HT uses the ring dimension of $2^{14}$ and hence bootstraps half as many bits per ciphertext as the v1.6.1 configuration. These speedups are the combined effect of the polynomial evaluation algorithm of Section~\ref{sec:ps}, the bootstrapping optimizations of Section~\ref{sec:algopt}, the parameter selection described above, and CPU-implementation-specific optimizations.

\FloatBarrier
\subsection{Continuous-CKKS Functional Bootstrapping}
\label{sec:results-continuous}

We evaluate continuous (Fourier-extension) functional bootstrapping for four functions: the identity, which corresponds to regular CKKS bootstrapping (Section~\ref{sec:prelim-continuous}), $\tanh$, sigmoid, and GELU. Three settings are considered: a low-latency setting with one level available after functional bootstrapping (Table~\ref{tab:continuous-main}), a setting with at least 10 levels at $N = 2^{16}$ (Table~\ref{tab:continuous-main-16}), and a higher-precision setting at $N = 2^{17}$ (Table~\ref{tab:continuous-main-17}). As in the discrete case, each table includes the multi-threaded CPU latency as a reference. We ran the experiments for both secret-key configurations; HT was as fast as SE or faster in all cases, so we do not provide any results for SE.

\paragraph{Low-latency setting.}
Table~\ref{tab:continuous-main} shows the results for the setting that targets the lowest latency: the scaling factor is 29 bits, which allows the use of 30-bit special primes and the ring dimension of $2^{15}$, and one multiplicative level is available between consecutive invocations of functional bootstrapping. For each function, GPU, and number of slots, we searched over the decomposition number $\beta$ and the level budgets for \textsc{CtS} and \textsc{StC}, and kept the parameter set with the smallest latency among those whose precision is within 0.3 bits of the best precision found for that configuration; the selected level budgets range from one level for each linear transformation for $n = 1$ to four or five levels for \textsc{CtS} and three levels for \textsc{StC} for $n \ge N/2$, and the selected $\beta$ is between 2 and 5. The measured precision is 7.4--7.5 bits for $n \ge N/2$ and 8.3--9.0 bits for 64 and 1024 slots; for a single slot, it varies more between functions and GPUs, from 6.8 to 11.4 bits.

Regular CKKS bootstrapping (the identity function) of a single slot takes 2.42 ms on RTX PRO 6000, and its latency stays below 5 ms for up to $N/2 = 16{,}384$ real values (4.82 ms); a ciphertext with $N/2$ complex slots, i.e., $N$ real values, is bootstrapped in 6.10 ms. On RTX 4090 and B200, the latency is below 5 ms for up to 1024 slots and reaches 5.90 ms and 5.14 ms, respectively, for $N/2$ real values. Evaluating a non-trivial function adds only 8\% to 26\% to the latency of the identity function, e.g., $\tanh$, sigmoid, and GELU are evaluated on 16,384 real values in about 5.5 ms on RTX PRO 6000. As in the discrete case, the latency grows slowly with the number of slots: it is 2.4x to 2.7x higher for $n = N$ than for $n = 1$. The GPU implementation is 54x to 78x faster than the multi-threaded CPU one in this setting.

These results can be directly compared with the results of Ouyang et al.~\cite{OLG+26}, which were obtained for the same number of available levels on the same GPU (RTX PRO 6000); the two sets of results are listed side by side in Table~\ref{tab:sota-continuous}. Their regular CKKS bootstrapping takes 8.2 ms for 4096 complex slots at about 5 bits of precision, whereas ours takes 6.10 ms for 16,384 complex slots at 7.4 bits of precision, i.e., we achieve a 1.3x lower latency for a ciphertext that carries four times as many slots. For continuous functional bootstrapping, the latencies of~\cite{OLG+26} are 24.5 ms for sigmoid and 40.4 ms for $\tanh$ and GELU for 4096 values at 6.6--9.2 bits of precision, whereas ours are about 5.5 ms for 16,384 real values at 7.4 bits of precision, i.e., our latency is 4.5x to 7.4x lower.

\begin{table}[!htbp]
  \centering
  \caption{Latency (``ms'') and measured output precision (``bits'') of continuous-CKKS functional bootstrapping (Fourier extension) of a single ciphertext with $n$ slots for the low-latency setting: $N = 2^{15}$, a 29-bit scaling factor, and one multiplicative level available between consecutive invocations of functional bootstrapping. The CPU results are for 36 threads, and the GPU results are for a single GPU. The identity function corresponds to regular CKKS bootstrapping. The \textsc{Hybrid\_Ternary} secret-key configuration and the \textsc{FlexibleAuto} scaling technique are used.}
  \label{tab:continuous-main}
  \small
  \setlength{\tabcolsep}{4.5pt}
  \begin{tabular}{ll rr rr rr rr rr}
    \toprule
    & & \multicolumn{2}{c}{$n = 1$} & \multicolumn{2}{c}{$n = 64$} & \multicolumn{2}{c}{$n = 1024$} & \multicolumn{2}{c}{$n = N/2$} & \multicolumn{2}{c}{$n = N$} \\
    \cmidrule(lr){3-4} \cmidrule(lr){5-6} \cmidrule(lr){7-8} \cmidrule(lr){9-10} \cmidrule(lr){11-12}
    Platform & Function & ms & bits & ms & bits & ms & bits & ms & bits & ms & bits \\
    \midrule
    Xeon 8360Y & Identity & 187 & 8.3 & 229 & 8.3 & 278 & 8.4 & 323 & 7.4 & 423 & 7.4 \\
     & $\tanh$ & 203 & 8.7 & 252 & 8.7 & 315 & 8.7 & 364 & 7.5 & 490 & 7.5 \\
     & Sigmoid & 202 & 8.7 & 268 & 8.7 & 325 & 8.7 & 366 & 7.5 & 511 & 7.5 \\
     & GELU & 207 & 8.8 & 284 & 8.8 & 334 & 8.8 & 369 & 7.5 & 501 & 7.5 \\
    \midrule
    RTX 4090 & Identity & 2.82 & 7.1 & 3.65 & 8.6 & 4.82 & 8.4 & 5.90 & 7.4 & 7.37 & 7.4 \\
     & $\tanh$ & 3.23 & 9.5 & 4.18 & 8.6 & 5.29 & 8.6 & 6.63 & 7.4 & 8.26 & 7.4 \\
     & Sigmoid & 3.23 & 9.5 & 4.26 & 8.5 & 5.43 & 8.6 & 6.79 & 7.4 & 8.80 & 7.5 \\
     & GELU & 3.28 & 11.1 & 4.43 & 8.8 & 5.70 & 8.7 & 6.88 & 7.4 & 8.63 & 7.5 \\
    \midrule
    RTX PRO 6000 & Identity & 2.42 & 8.2 & 3.08 & 8.3 & 4.14 & 8.3 & 4.82 & 7.4 & 6.10 & 7.4 \\
     & $\tanh$ & 2.61 & 9.5 & 3.57 & 8.6 & 4.53 & 8.7 & 5.49 & 7.4 & 7.01 & 7.4 \\
     & Sigmoid & 2.69 & 9.2 & 3.69 & 8.5 & 4.58 & 8.7 & 5.48 & 7.4 & 7.02 & 7.5 \\
     & GELU & 2.71 & 11.2 & 3.79 & 9.0 & 4.84 & 8.7 & 5.55 & 7.4 & 7.36 & 7.5 \\
    \midrule
    B200 & Identity & 2.79 & 6.8 & 3.22 & 8.3 & 4.29 & 8.3 & 5.14 & 7.4 & 6.61 & 7.4 \\
     & $\tanh$ & 3.10 & 9.6 & 3.68 & 8.7 & 4.87 & 8.8 & 5.74 & 7.4 & 7.54 & 7.4 \\
     & Sigmoid & 3.11 & 9.7 & 3.95 & 8.9 & 4.84 & 8.7 & 5.75 & 7.4 & 7.63 & 7.5 \\
     & GELU & 3.17 & 11.4 & 4.06 & 8.9 & 5.08 & 8.8 & 5.83 & 7.4 & 7.72 & 7.5 \\
    \bottomrule
  \end{tabular}
\end{table}

\paragraph{Parameter selection for the settings with more levels.}
Tables~\ref{tab:continuous-main-16} and~\ref{tab:continuous-main-17} show the results for fully packed ciphertexts when multiple levels are available after functional bootstrapping. Both the case of complex inputs, where all $N/2$ complex slots are used (and can encode $N$ real values), and the case of real inputs, where $N/2$ real values are processed, are considered. For $N = 2^{16}$, we require at least 10 levels after functional bootstrapping and use the same scaling factor for all configurations of Table~\ref{tab:continuous-main-16}: 40 bits, which is the largest scaling factor that leaves at least 10 levels for every function and every $\beta \in \{3, 4, 5\}$. For each function, GPU, and $\beta$, we then searched over the level budgets (up to five levels for \textsc{CtS} and four levels for \textsc{StC}) and the size of the special primes (30 or 60 bits), kept the parameter set with the smallest latency, and report the number of levels that this parameter set leaves after functional bootstrapping. For $N = 2^{17}$ (Table~\ref{tab:continuous-main-17}), the total number of RNS limbs is the binding constraint, as FIDESlib supports at most 64 limbs. For $\beta = 3$, we first increased the number of levels as far as the limbs of $Q$ and $P$ stay within this limit, and then increased the scaling factor until $\log (Q_L P)$ reached the 128-bit security bound for $N = 2^{17}$. This yields a 54-bit scaling factor, 60-bit special primes, level budgets of five levels for \textsc{CtS} and four levels for \textsc{StC}, and 63 RNS limbs in total (48 for $Q$ and 15 for $P$).

\paragraph{Results for $N = 2^{16}$.}
With a 40-bit scaling factor, the measured precision is 16.7--16.8 bits, and 10 levels are available after functional bootstrapping (11 levels in a few cases for $\beta = 5$). For complex inputs, the lowest latency is 34.8 ms for the identity function and about 39 ms for the other functions on B200; the corresponding latencies on RTX PRO 6000 are 38.4 ms and about 43 ms, and RTX 4090 is 1.5x to 1.6x slower than RTX PRO 6000. In contrast to the low-latency setting, B200 is the fastest GPU for these larger parameters. For real inputs, the latency is about 30\% lower, e.g., 24.2 ms for the identity function and 26.6 ms for $\tanh$ and sigmoid on B200. The decomposition numbers of 3 and 4 result in similar latencies on the GPUs (within 7\% of each other), while $\beta = 5$ is slower than the better of the two in all cases. The GPU implementation is 26x to 60x faster than the multi-threaded CPU one.

\begin{table}[!htbp]
  \centering
  \caption{Continuous-CKKS functional bootstrapping (Fourier extension) of a single fully packed ciphertext for $N = 2^{16}$, a 40-bit scaling factor, and decomposition numbers $\beta \in \{3, 4, 5\}$, with at least 10 levels available after functional bootstrapping; ``Lvl'' is the number of levels available after functional bootstrapping, and ``ms'' is the latency. The measured output precision is 16.7 to 16.8 bits in all configurations. The upper part is for complex inputs ($N/2$ complex slots, i.e., $N$ real values), and the lower part is for real inputs ($N/2$ real values). The CPU results are for 36 threads, and the GPU results are for a single GPU. The \textsc{Hybrid\_Ternary} secret-key configuration and the \textsc{FlexibleAuto} scaling technique are used.}
  \label{tab:continuous-main-16}
  \small
  \begin{tabular}{ll rr rr rr}
    \toprule
    & & \multicolumn{2}{c}{$\beta = 3$} & \multicolumn{2}{c}{$\beta = 4$} & \multicolumn{2}{c}{$\beta = 5$} \\
    \cmidrule(lr){3-4} \cmidrule(lr){5-6} \cmidrule(lr){7-8}
    Platform & Function & Lvl & ms & Lvl & ms & Lvl & ms \\
    \midrule
    \multicolumn{8}{l}{Complex inputs ($N/2$ complex slots)} \\
    \midrule
    Xeon 8360Y & Identity & 10 & 1822 & 10 & 1640 & 11 & 1809 \\
     & $\tanh$ & 10 & 2291 & 10 & 1825 & 10 & 1997 \\
     & Sigmoid & 10 & 2366 & 10 & 1830 & 11 & 2025 \\
     & GELU & 10 & 2441 & 10 & 1905 & 10 & 1989 \\
    \addlinespace
    RTX 4090 & Identity & 10 & 60.1 & 10 & 60.0 & 11 & 69.4 \\
     & $\tanh$ & 10 & 67.2 & 10 & 67.8 & 10 & 71.4 \\
     & Sigmoid & 10 & 67.3 & 10 & 67.7 & 11 & 71.9 \\
     & GELU & 10 & 68.3 & 10 & 68.7 & 10 & 72.5 \\
    \addlinespace
    RTX PRO 6000 & Identity & 10 & 38.7 & 10 & 38.4 & 11 & 44.0 \\
     & $\tanh$ & 10 & 43.2 & 10 & 43.3 & 10 & 45.4 \\
     & Sigmoid & 10 & 43.2 & 10 & 43.3 & 10 & 45.5 \\
     & GELU & 10 & 43.8 & 10 & 44.1 & 10 & 46.1 \\
    \addlinespace
    B200 & Identity & 10 & 34.8 & 10 & 35.3 & 11 & 39.5 \\
     & $\tanh$ & 10 & 40.2 & 10 & 38.5 & 10 & 41.0 \\
     & Sigmoid & 10 & 40.2 & 10 & 38.4 & 10 & 41.0 \\
     & GELU & 10 & 40.7 & 10 & 39.1 & 10 & 41.9 \\
    \midrule
    \multicolumn{8}{l}{Real inputs ($N/2$ real values)} \\
    \midrule
    Xeon 8360Y & Identity & 10 & 1292 & 10 & 1183 & 11 & 1284 \\
     & $\tanh$ & 10 & 1602 & 10 & 1292 & 10 & 1353 \\
     & Sigmoid & 10 & 1596 & 10 & 1280 & 11 & 1325 \\
     & GELU & 10 & 1708 & 10 & 1294 & 10 & 1314 \\
    \addlinespace
    RTX 4090 & Identity & 10 & 41.9 & 10 & 41.7 & 11 & 47.3 \\
     & $\tanh$ & 10 & 47.0 & 10 & 47.2 & 10 & 48.3 \\
     & Sigmoid & 10 & 47.0 & 10 & 47.1 & 10 & 48.3 \\
     & GELU & 10 & 47.6 & 10 & 47.8 & 10 & 49.0 \\
    \addlinespace
    RTX PRO 6000 & Identity & 10 & 27.7 & 10 & 27.4 & 11 & 30.8 \\
     & $\tanh$ & 10 & 31.2 & 10 & 31.0 & 10 & 31.4 \\
     & Sigmoid & 10 & 31.2 & 10 & 30.9 & 10 & 31.4 \\
     & GELU & 10 & 31.5 & 10 & 31.3 & 10 & 31.9 \\
    \addlinespace
    B200 & Identity & 10 & 24.2 & 10 & 24.7 & 11 & 27.1 \\
     & $\tanh$ & 10 & 28.4 & 10 & 26.6 & 10 & 27.9 \\
     & Sigmoid & 10 & 28.5 & 10 & 26.6 & 10 & 27.9 \\
     & GELU & 10 & 28.7 & 10 & 26.8 & 10 & 28.5 \\
    \bottomrule
  \end{tabular}
\end{table}

\paragraph{Higher-precision setting.}
For $N = 2^{17}$, functional bootstrapping achieves 28.5 bits of precision and leaves 24 levels for the identity function and 23 levels for the other functions. On B200, a fully packed ciphertext is bootstrapped in 129.2 ms for the identity function and in 135--138 ms for the other functions in the case of complex inputs, and in 85.0 ms and 88--89 ms, respectively, in the case of real inputs; RTX PRO 6000 is about 15\% slower. In this setting, the choice of the function has a small effect on the latency (at most 7\%), and the GPU implementation is 40x to 49x faster than the multi-threaded CPU one. The results for RTX 4090 are not available because its GPU memory is insufficient for these configurations.

\begin{table}[!htbp]
  \centering
  \caption{Continuous-CKKS functional bootstrapping (Fourier extension) of a single fully packed ciphertext for the higher-precision setting: $N = 2^{17}$, a 54-bit scaling factor, and $\beta = 3$, for complex inputs ($N/2$ complex slots, i.e., $N$ real values) and real inputs ($N/2$ real values); ``Lvl'' is the number of levels available after functional bootstrapping (the same for complex and real inputs). The measured output precision is 28.5 bits in all configurations. The CPU results are for 36 threads, and the GPU results are for a single GPU. The \textsc{Hybrid\_Ternary} secret-key configuration and the \textsc{FlexibleAuto} scaling technique are used.}
  \label{tab:continuous-main-17}
  \small
  \begin{tabular}{ll r rr}
    \toprule
    & & & \multicolumn{2}{c}{Latency [ms]} \\
    \cmidrule(lr){4-5}
    Platform & Function & Lvl & Complex inputs & Real inputs \\
    \midrule
    Xeon 8360Y & Identity & 24 & 5922 & 4180 \\
     & $\tanh$ & 23 & 6168 & 4287 \\
     & Sigmoid & 23 & 6140 & 4269 \\
     & GELU & 23 & 6232 & 4316 \\
    \midrule
    RTX PRO 6000 & Identity & 24 & 147.9 & 98.8 \\
     & $\tanh$ & 23 & 153.8 & 101.6 \\
     & Sigmoid & 23 & 153.8 & 101.6 \\
     & GELU & 23 & 156.7 & 103.1 \\
    \midrule
    B200 & Identity & 24 & 129.2 & 85.0 \\
     & $\tanh$ & 23 & 135.0 & 87.8 \\
     & Sigmoid & 23 & 135.0 & 87.8 \\
     & GELU & 23 & 137.9 & 89.3 \\
    \bottomrule
  \end{tabular}
\end{table}

\FloatBarrier
\section{Concluding Remarks}
\label{sec:conclusions}

We presented CPU and GPU implementations of discrete-CKKS and continuous-CKKS functional bootstrapping that combine a redesigned Paterson--Stockmeyer evaluation of large-degree polynomials with several optimizations that reduce the multiplicative depth and the computational cost of bootstrapping. Our results show that the standard CKKS bootstrapping workflow, which is based on a polynomial approximation of the modular reduction and of its functional generalizations, can reach the latencies that were previously associated with blind-rotation-based designs, provided that the evaluation of large-degree polynomials and the multiplicative depth of the bootstrapping circuit are carefully optimized. In contrast to these designs, the same workflow extends to fully packed ciphertexts and to large ring dimensions, i.e., to a higher throughput, a higher precision, and more levels available after bootstrapping. In practical terms, this changes the criteria for choosing between CKKS and CGGI for the evaluation of lookup tables over small integers: on consumer-grade and professional GPUs, CKKS functional bootstrapping is now competitive with CGGI programmable bootstrapping not only in throughput but also in latency, even when only a handful of values have to be bootstrapped, and it additionally supports the evaluation of functions over real numbers. We intend to make our implementation available in later releases of OpenFHE and FIDESlib.

\section*{Acknowledgments}
A. Al Badawi, A. Alexandru, C. Pascoe, and Y. Polyakov were funded, in
part, by the Advanced Research Projects Agency for Health (ARPA-H). The views
and conclusions contained in this document are those of the authors and should
not be interpreted as representing the official policies, either expressed or implied, of the U.S. Government. C. Agulló is supported by Grant 23039/GERM/25, funded by FSRM (10.13039/100007801). The authors acknowledge the use of LLMs for proofreading and for instantiating scripts for experiments.
The authors have reviewed and verified all generated content and assume full
responsibility for the final manuscript.

\bibliographystyle{alpha}
\bibliography{references}

\end{document}